\documentclass{aastex63}

\usepackage{amsmath}
\usepackage{units}
\usepackage{physics}

\begin{document}

\title{Physical Parameters of Late-type Contact Binaries in the Northern Catalina Sky Survey}

\correspondingauthor{Weijia Sun, Xiaodian Chen}
\email{swj1442291549@gmail.com, chenxiaodian@nao.cas.cn}

\author[0000-0002-3279-0233]{Weijia Sun}
\affiliation{Department of Astronomy, School of Physics, Peking University, Beijing 100871, China}
\affiliation{Key Laboratory for Optical Astronomy, National Astronomical Observatories, Chinese Academy of Sciences, 20A Datun Road, Chaoyang District, Beijing 100012, China}
\affiliation{Department of Physics and Astronomy, Macquarie University, Balaclava Road, Sydney, NSW 2109, Australia}
\affiliation{Centre for Astronomy, Astrophysics and Astrophotonics, Macquarie University, Balaclava Road, Sydney, NSW 2109, Australia}

\author[0000-0001-7084-0484]{Xiaodian Chen}
\affiliation{Key Laboratory for Optical Astronomy, National Astronomical Observatories, Chinese Academy of Sciences, 20A Datun Road, Chaoyang District, Beijing 100012, China}
\affiliation{School of Astronomy and Space Science, University of the Chinese Academy of Sciences, Huairou 101408, China}
\affiliation{Department of Astronomy, China West Normal University, Nanchong 637002, China}

\author[0000-0001-9073-9914]{Licai Deng}
\affiliation{Key Laboratory for Optical Astronomy, National Astronomical Observatories, Chinese Academy of Sciences, 20A Datun Road, Chaoyang District, Beijing 100012, China}
\affiliation{School of Astronomy and Space Science, University of the Chinese Academy of Sciences, Huairou 101408, China}
\affiliation{Department of Astronomy, China West Normal University, Nanchong 637002, China}

\author[0000-0002-7203-5996]{Richard de Grijs}
\affiliation{Department of Physics and Astronomy, Macquarie University, Balaclava Road, Sydney, NSW 2109, Australia}
\affiliation{Centre for Astronomy, Astrophysics and Astrophotonics, Macquarie University, Balaclava Road, Sydney, NSW 2109, Australia}
\affiliation{International Space Science Institute--Beijing, 1 Nanertiao, Hai Dian District, Beijing 100190, China}

\begin{abstract} 
We present the physical parameters of 2335 late-type contact binary (CB) systems extracted from the Catalina Sky Survey (CSS). Our sample was selected from the CSS Data Release 1 by strictly limiting the prevailing temperature uncertainties and light-curve fitting residuals, allowing us to almost eliminate any possible contaminants. We developed an automatic Wilson--Devinney-type code to derive the relative properties of CBs based on their light-curve morphology. By adopting the distances derived from CB (orbital) period--luminosity relations (PLRs), combined with the well-defined mass--luminosity relation for the systems' primary stars and assuming solar metallicity, we calculated the objects' masses, radii, and luminosities. Our sample of fully eclipsing CBs contains 1530 W-, 710 A-, and 95 B-type CBs. A comparison with literature data and with the results from different surveys confirms the accuracy and coherence of our measurements. The period distributions of the various CB subtypes are different, hinting at a possible evolutionary sequence. W-type CBs are clearly located in a strip in the total mass versus mass ratio plane, while A-type CBs may exhibit a slightly different dependence. There are no significant differences among the PLRs of A- and W-type CBs, but the PLR zero points are affected by their mass ratios and fill-out factors. Determination of zero-point differences for different types of CBs may help us improve the accuracy of the resulting PLRs. We demonstrate that automated approaches to deriving CB properties could be a powerful tool for application to the much larger CB samples expected to result from future surveys.
\end{abstract}


\section{Introduction}

Late-type contact binary systems (CBs), also known as W Ursae Majoris (W UMa) variables, are eclipsing binaries where both components fill their Roche lobes. Hence, they are in `contact' with each other, thus allowing mass and energy transfer \citep{1968ApJ...153..877L}. The components' close separation facilitates relatively short orbital periods, with most systems having periods between 0.25 and 0.5 days. Another natural outcome of their proximity is the variability of their light curves. The latter are effective tools to study CB formation and evolution.

Previous studies have revealed that CBs are embedded in a common envelope \citep{1979ApJ...231..502L} with both components having similar temperatures \citep{1941ApJ....93..133K}, although the systems may undergo periodic thermal-relaxation oscillations \citep{1976ApJ...205..217F, 1977MNRAS.179..359R}. However, an unresolved mystery remains as to whether an evolutionary sequence exists among different types of CBs. Only limited sample sizes, encompassing just tens of CBs with common characteristics, have thus far been available for comparative research \citep[e.g.,][]{2001MNRAS.328..635Q, 2005ApJ...629.1055Y, 2013MNRAS.430.2029Y}. The large sample size is essential to constrain evolutionary models of CBs \citep{2006AcA....56..199S}, as well as their angular-momentum loss properties and nuclear evolutionary pathways, particularly as regards any impact these may have on the resulting orbital periods \citep{2016ApJ...832..138C, 2019MNRAS.tmp.3207J} and the evolutionary products of the different CB types \citep{2015AJ....150...69Y, 2019MNRAS.485.4588L}.

Since \citet{1967MmRAS..70..111E} first proposed to use CBs as distance indicators, various studies have attempted to establish period--luminosity (PL)--color (PLC) relations \citep{1994PASP..106..462R, 2016AJ....152..129C}. \citet{2018ApJ...859..140C} managed to achieve a distance accuracy of 7\% using infrared passbands\footnote{This was improved to 6\% based on \textit{Gaia} Data Release 2 measurements \citep{2019gaia.confE..60C}.}. This may be further improved if we can exclude the possible impact associated with using different subtypes and any dependence on the CBs' physical parameters. However, this will only be feasible based on large sample sizes.

The sample of known CBs was recently significantly increased thanks to new data from several sky surveys that provide high-cadence, long-term, high-precision photometric observations in a range of passbands, including, e.g., the Catalina Sky Survey \citep[CSS;][]{2017MNRAS.465.4678M}, the \textit{Wide-field Infrared Survey Explorer} catalogue \citep[\textit{WISE};][]{2018ApJS..237...28C}, the All-Sky Automated Survey for Supernovae \citep[ASAS-SN;][]{2018MNRAS.477.3145J}, the Northern Sky Variability Survey \citep[NSVS;][]{2006AJ....131..621G}, and the Asteroid Terrestrial-impact Last Alert System \citep[ATLAS;][]{2018AJ....156..241H}. As sample sizes increased, researchers have taken advantage of the data from various surveys and constructed genuine CB samples for further statistical study \citep{1995ApJ...446L..19R, 2011A&A...528A..90N, 2017MNRAS.465.4678M}. However, most previous studies dealing with large samples of CBs were limited to analyses of their light-curve morphology (e.g., periods and amplitudes), which is rather different from deriving the intrinsic properties of the stellar components. Moreover, future surveys using, e.g., the Zwicky Transient Facility \citep[ZTF;][]{2019PASP..131a8002B} and the Large Synoptic Survey Telescope \citep[LSST;][]{2009arXiv0912.0201L} will likely result in enormous numbers of newly discovered CBs, thus posing a challenge to our ability to derive stellar parameters based on individual light-curve solutions.

In this paper, we develop an automated Wilson--Devinney-type \citep[W--D;][]{1971ApJ...166..605W, 1979ApJ...234.1054W} code to derive physical parameters from the CB light curves, and we apply our method to a large CB sample from the CSS Data Release 1 \citep[CSDR1\footnote{\url{http://nesssi.cacr.caltech.edu/DataRelease/}};][]{2014ApJS..213....9D}. Armed with distance information obtained from PLR analysis in infrared passbands \citep{2018ApJ...859..140C}, we can estimate the intrinsic properties---masses, radii, and luminosities---of 2335 CBs.

This article is organized as follows. In Section~\ref{sec:data}, we describe the data and candidate selection. The details of the method and the input parameters, as well as the selection criteria applied to obtain our final catalog, are discussed in Section~\ref{sec:lc}. We performed a series of tests to verify the accuracy and consistency of our measurements, which we report in Section~\ref{sec:validation}. Section~\ref{sec:discussion} presents a discussion of the CB-subtype classification, their evolutionary states, and implications for the PLRs, which is followed by a summary in Section~\ref{sec:conclusions}.

\section{Data and Candidate Selection \label{sec:data}}

We used CB data from the CSDR1, the northern-sky section of the CSS. The survey used three telescopes to cover the sky between declinations $\delta = -75\degr$ and $+70\degr$ at Galactic latitudes $\left|b\right|> 15\degr$. The unfiltered observations were transformed to $V_\mathrm{CSS}$ magnitudes \citep{2013ApJ...763...32D}. The CSDR1 collected $\sim 47,000$ periodic variables based on their analysis of 5.4 million variable star candidates, with a median number of observations per candidate system of around 250. Because of limitations to the aperture photometry obtained, the $V$-band zero-point uncertainty is $\sim\unit[0.06-0.08]{mag}$ from field to field. The photometric uncertainties were determined by employing an empirical relationship between the source fluxes and the observed photometric scatter. Typical values range from 0.05 to $\unit[0.10]{mag}$, mainly depending on the target brightness.

The initial CB sample was selected as described by \citet{2014ApJS..213....9D}. Based on the Stetson variability index ($J_{WS}$) and its standard deviation ($\sigma_J$), the authors selected a sample of variable stars from the reduced photometric data. For classification purposes, a Lomb–-Scargle-type \citep{1976Ap&SS..39..447L, 1982ApJ...263..835S} periodogram analysis was applied to all variable candidates. Those with significant periodic patterns were subsequently studied using the Adaptive Fourier Decomposition method \citep{2015MNRAS.446.2251T} to derive their best-fitting periods. Finally, the remaining candidates were visually inspected and classified based on their periods, light-curve morphologies, and colors.

\citet{2014ApJS..213....9D} found 30,743 CBs (EW-type stars) in the CSDR1. To estimate their temperatures from multi-band photometry, we cross-matched the sample with the American Association of Variable Star Observers' (AAVSO) Photometric All-Sky Survey \citep[APASS;][]{2014CoSka..43..518H}. This is a survey in the $B, V$, and Sloan $g^\prime, r^\prime$, and $i^\prime$ passbands. Its Data Release (DR) 9 covers almost the entire sky \citep{2016yCat.2336....0H} and provides high-accuracy APASS photometry without any offsets \citep{2014AJ....148...81M}. Following cross-matching, we found 13,726 CB candidates for which both CSDR1 and APASS photometry had been obtained. Comparison with the LINEAR data of \citet{2013AJ....146..101P}, for which \citet{2014ApJS..213....9D} found that 98.3\% of CBs had the same classification, suggests only a minor contribution from contaminants. Given that the candidates used in our subsequent analysis comprise a subset of the initial sample (candidates with poor mass-ratio determinations or low inclinations were ignored; see Section~\ref{sec:qsearch}), we also expect a low to a negligible level of contamination in our CB sample.

\section{light-curve solutions \label{sec:lc}}

To model the W UMa light curves, we used a W--D-type approach. Our program executes two subroutines, one for generating light and radial velocity curves based on a given set of physical parameters and the other allowing adjustments of the light- and velocity-curve parameters using differential corrections. We adopted `Mode 3,' appropriate for over-contact binaries, with both component stars filling their Roche lobes. The component stars can still have different surface brightnesses if they are in geometric contact without being in thermal equilibrium.

\subsection{Effective temperatures\label{sec:temp}}

The effective temperature is one of the W--D code's primary input parameters. Light curve morphologies can place tight constraints on the $T_2/T_1$ temperature ratio, but not on the individual component temperatures. Therefore, we estimated the effective temperatures based on the CB's spectral type, as inferred from its intrinsic color, using the de-reddened $(B-V)_0$ APASS photometry.

We adopted the relevant $E(B-V)$ reddening values from the 3D dust extinction map derived from Pan-STARRS1 and 2MASS photometry by \citep{2019ApJ...887...93G}. Distances to our sample CBs were obtained on the basis of the \citet{2018ApJ...859..140C} PLRs for 12 optical to mid-infrared bands based on 183 nearby W UMa-type CBs with accurate \textit{Tycho}--\textit{Gaia} parallaxes. These authors determined the distances to field CBs by combining the PLR distances based on {\textit WISE}/$W_1$, \textit{Gaia}/$G_\mathrm{mean}$ (DR 1), and Two Micron All-Sky Survey (2MASS)/$JHK_\mathrm{s}$ photometry \citep[][their Section 5.2]{2018ApJ...859..140C}.

The reddening in the $B$ and $V$ passbands was calculated by employing $A_\lambda/E(B-V)$ coefficients from \citet[][their Table 6]{2011ApJ...737..103S}, for $R_V = 3.1$; here, $A_\lambda$ denotes the extinction in a given bandpass $\lambda$. The median $E(B-V)$ value is \unit[0.037]{mag}, while 90\% of our sample objects have reddening values lower than \unit[0.15]{mag}.

We then used the empirical relation between the intrinsic color, $(B-V)_0$, and the average temperature, $T$, from \citet{2013ApJS..208....9P} to estimate the color temperature, $T_\mathrm{color}$. This approximate estimation is sufficient, since it only affects the determination of the absolute temperatures, while it has a minor effect on other key parameters, including the mass ratio, relative radii, and the system's inclination. To better illustrate this, we compared the temperatures derived here with those obtained from a low-resolution spectroscopic survey undertaken with the Large Sky Area Multi-Object Fiber Spectroscopic Telescope \citep[LAMOST;][]{2015RAA....15.1095L}. LAMOST \citep{2012RAA....12.1197C, 2012RAA....12..735D, 2012RAA....12..723Z} is a reflective Schmidt telescope located at Xinglong Observatory north of Beijing, China, with an effective aperture of \unit[3.6-4.9]{m} and a field of view of $5\arcdeg$ (diameter). It has 4000 fibers covering its focal plane. Its wavelength coverage is $\unit[3650-9000]{\AA}$, with a spectral resolution of $R\sim 1800$. LAMOST is an effective facility to study the physical properties of binary systems \citep[e.g.,][]{2017RAA....17...87Q}.

\begin{figure}[ht!]
\plotone{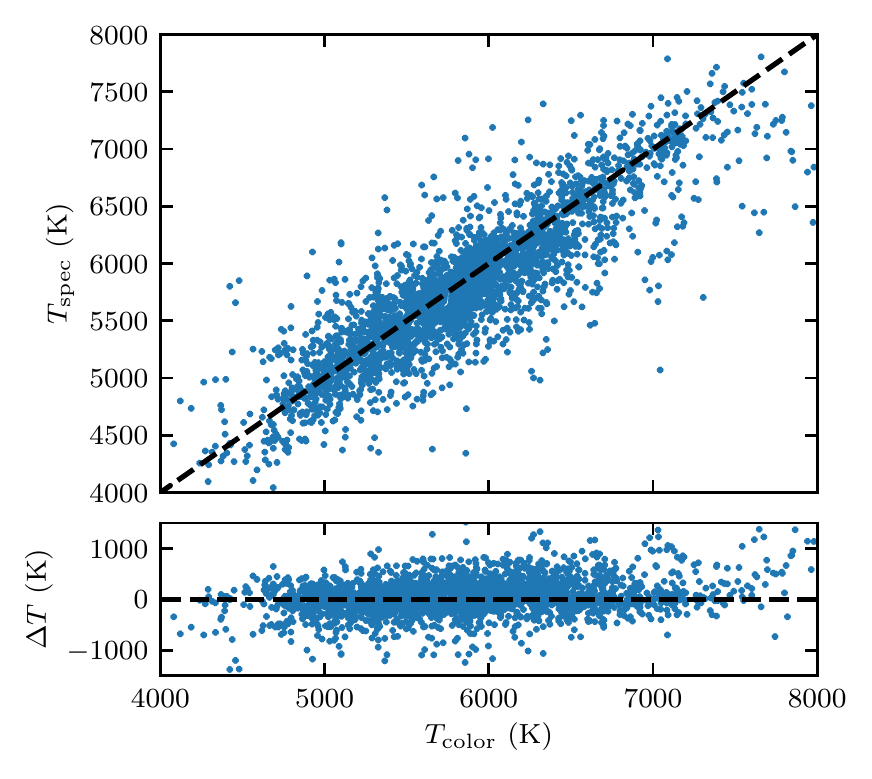}
\caption{Temperature measurements based on intrinsic colors $(B-V)_0$ and spectroscopic data. The black dashed line is the one-to-one linear relation. The residual temperature, $\Delta T = T_\mathrm{color} - T_\mathrm{spec}$, is displayed in the bottom panel. The root-mean-square error (RMSE) is \unit[352]{K}. \label{fig:teff}}
\end{figure}

By cross-matching our CB sample with the LAMOST DR 5 catalog of A-, F-, G-, and K-type stars, we found that the LAMOST survey has collected spectra of 2930 of our sample stars. In Fig.~\ref{fig:teff}, we present our temperature measurements based on color ($T_\mathrm{color}$) and spectroscopic data ($T_\mathrm{spec}$), as well as the residual, $\Delta T = T_\mathrm{color} - T_\mathrm{spec}$. There is no significant bias apparent toward any temperature. The root-mean-square error (RMSE) is \unit[352]{K}, which is close to the mean error in the temperature determination (\unit[324]{K}) for CBs derived from SDSS colors \citep{2017MNRAS.465.4678M}.

The color index is commonly used as a proxy for the temperature of the primary component. However, this approximation will introduce biases in temperature for both components. To alleviate this problem, we assigned the color temperature to the system's combined light rather than just to the primary star. We hence introduce the combined temperature, $T_\mathrm{c}$, as
\begin{equation}
{T_c}^4 = \frac{L_\mathrm{p} + L_\mathrm{s}}{L_\mathrm{p}/{T_\mathrm{p}}^4+L_\mathrm{s}/{T_\mathrm{s}}^4} \label{eq:comb}
\end{equation}
where $L_\mathrm{p}$ ($T_\mathrm{p}$) and $L_\mathrm{s}$ ($T_\mathrm{s}$) are the luminosities (temperatures) of the primary and secondary components, respectively. In practice, we adopted $T_\mathrm{color}$ for the primary star's temperature in the first run, and we then obtained the corresponding luminosities and temperatures for both components. Next, we calculated $T_\mathrm{c}$ and the ratio of $T_\mathrm{c}$ and $T_\mathrm{color}$, using Eq.~\ref{eq:comb} and $\alpha=T_\mathrm{c}/T_\mathrm{color}$, respectively. We subsequently corrected the individual temperatures by dividing them by $\alpha$. These new temperatures were taken as input for a second run, which yielded a new solution that retained the combined temperature, $T_\mathrm{c}$, close to the color temperature, $T_\mathrm{color}$.

\subsection{Other parameters}

\citet{2017MNRAS.465.4678M} found that a photospheric temperature of \unit[6200]{K} separates CBs into two groups. Systems with temperatures greater than \unit[6200]{K} generally have smaller amplitudes ($\lesssim \unit[0.5]{mag}$), while the amplitudes of cooler CBs can reach \unit[0.8]{mag}. This temperature corresponds to the transition between radiative and convective energy transport. Hotter main-sequence (MS) stars ($T>\unit[6200]{K}$) are dominated by radiative energy transport at the surface, while cooler MS stars have convective envelopes \citep{2012sse..book.....K}. Therefore, we adopt the relevant gravity-darkening exponents, $g = 0.32$ and $g = 1.0$ \citep{1980MNRAS.193...79R}, for convective and radiative energy transport, respectively. The corresponding bolometric albedos are $A = 0.5$ and $A = 1.0$, which is a reasonable approximation given that \citet{1980MNRAS.193...79R} found that the expected bolometric albedo for stars with radiative envelopes is 1.0, while the average value for those with convective envelopes is around 0.5. We adopted the logarithmic limb-darkening law of \citep{1970AJ.....75..175K}; its coefficients have been tabulated by \citep{1993AJ....106.2096V}.

Since we have no information about the metallicity of our CBs, we adopted solar metallicity. This is statistically acceptable since our sample is located within $\unit[2-3]{kpc}$ from the Sun. Meanwhile, we assume a zero rate of period change ($\dd P / \dd t = 0$), because uniform orbital period changes are unusual among CBs \citep[e.g.,][]{1977ivsw.conf..393K, 2001MNRAS.328..635Q}. Next, the dimensionless surface potential $\Omega$ was calculated using the formulation of \citet{1979ApJ...234.1054W}. Note that for over-contact binaries $\Omega_2$ is fixed to the same value as $\Omega_1$, and thus we used the same potential $\Omega_1$ for both the primary and secondary stars.

We did not consider the effect of starspots, for reasons of clarity and simplicity. This is a generally accepted practice since spots usually have only subtle effects on the shape of a light curve. Spots are usually included to explain asymmetries when one light-curve maximum is higher than the other, an effect also known as the O'Connell effect. As explained in Section~\ref{sec:qsearch}, we removed those solutions that did not fit the light curves well. Therefore, any CBs that are strongly affected by the O'Connell effect have already been excluded from our sample. We remind the reader that one should exercise caution in reaching the simplistic conclusion that our sample CBs may be free from spots because the hypothetical distribution of spots is by no means uniquely determined by the CB light curves. 

We also assumed that the third-light contribution is negligible. Any tertiary component does not affect the estimation of the relative parameters (including the mass ratios and inclinations) but only the luminosities and masses. \citet{2006AJ....132..650D} performed a spectroscopic search for third members in their sample of CB systems. They found that the uncertainty in total luminosity introduced by a tertiary component is smaller than \unit[0.15]{mag}, leading to an increase in the uncertainty in the derived masses of only $\sim3$\%.

\subsection{The $q$-search method \label{sec:qsearch}}

Using the periods derived by \citet{2014ApJS..213....9D}, we converted our light curves from the time domain to the phase domain. 

Next, we used Gaussian Process (GP) models to fit the photometric data and reject the outliers. GP modeling, which is well suited to time-series modeling, is routinely and widely applied to the light curves of transits \citep[e.g.,][]{2012MNRAS.419.2683G, 2012MNRAS.422..753G, 2013ApJ...772L..16E} and variable stars \citep[e.g.,][]{2012RSPTA.37110550R, 2017MNRAS.464.1353M}. For our purposes, we selected a GP kernel composed of a Mat{\'e}rn component and an amplitude factor, as well as observational noise. The Mat{\'e}rn kernel with $\nu=3/2$ was chosen for its great capability to recreate the light curves' features. We thus calculated the predicted light curve for a given object using the GP model and the corresponding posterior standard deviation ($\sigma$); $\unit[2]{\sigma}$ outliers were rejected to allow for a robust light-curve analysis.

\begin{figure*}[ht]
\plottwo{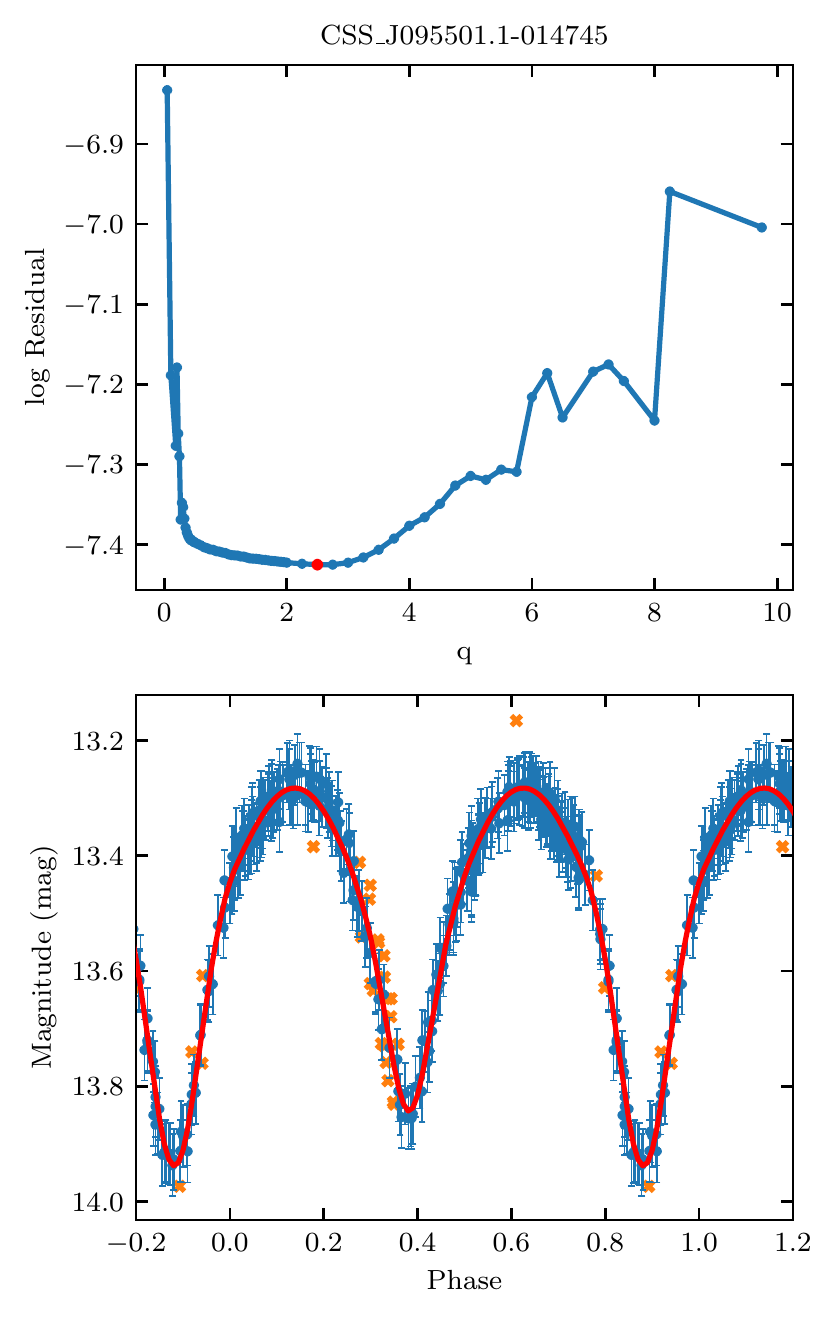}{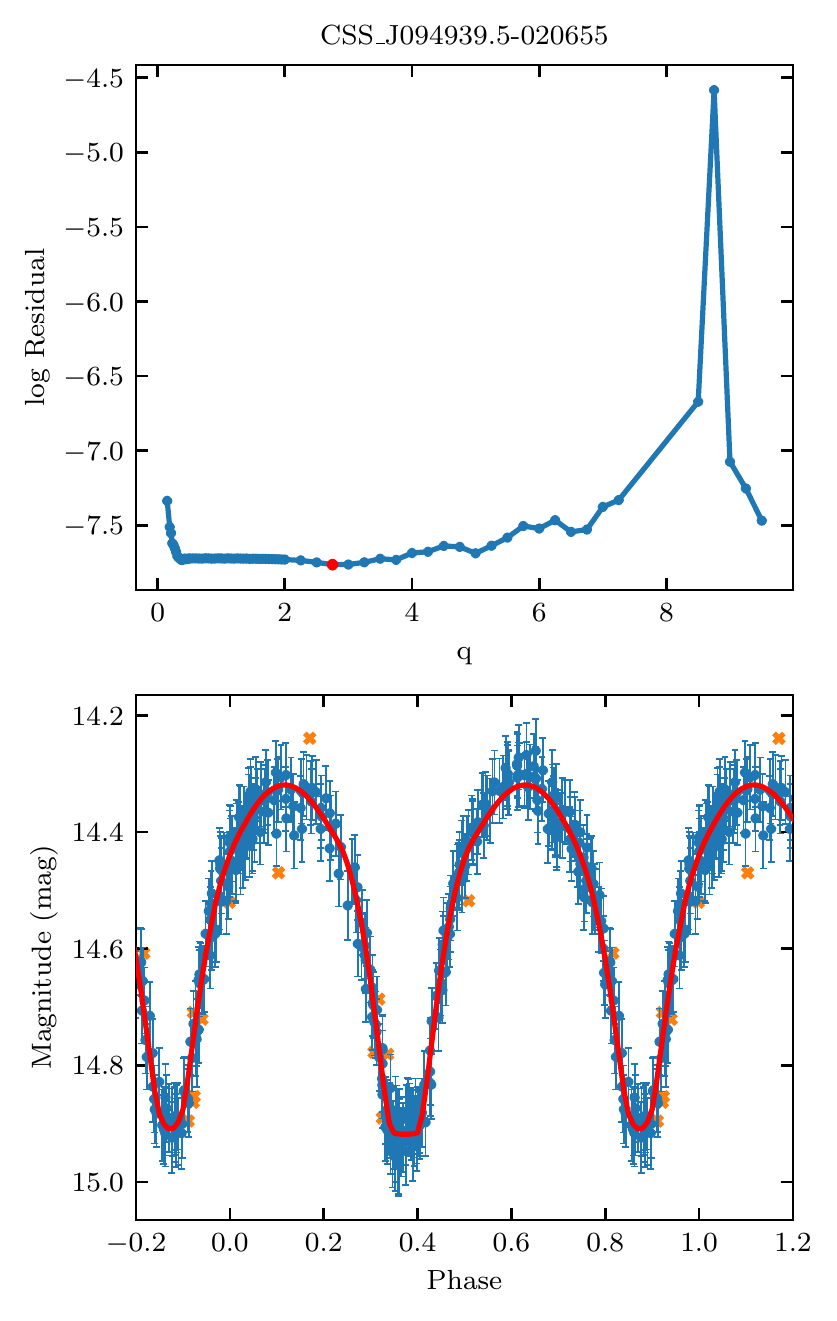}
\caption{Two examples of the $q$-search diagram. (top) Mean residual versus $q$. Red dots represent the best $q$ corresponding to the smallest residuals. (bottom) Observations and best-fitting solutions of the light curves. Blue dots show the observational data and their photometric errors, while orange crosses are outliers that were rejected through GP regression. The best solutions (red curves) were derived based on the $q$ values given in the first row. \label{fig:qdiagram}}
\end{figure*}

To constrain the CB mass ratios, $q=m_2/m_1$, we employed a $q$-search method, i.e., we analyzed how the mean residual changes for different, fixed $q$ values, adopting the $q$ value corresponding to the minimum mean residual as the best light-curve solution (see Fig.~\ref{fig:qdiagram}, top row). This is an effective approach to estimating CB mass ratios without having access to information pertaining to the radial velocity curves \citep{2005Ap&SS.296..221T}. It has been widely applied \citep[e.g.,][]{2016AJ....152..129C, 2017PASJ...69...69Y, 2018PASJ...70...87Z}. Next, we adopted the standard error given by the W--D code through the Method of Multiple Subsets (MMS) as the uncertainty associated with the relevant derived property (except for $q$; see Section~\ref{sec:validation}.).

To be more specific, we first fixed the value of $q$, leaving as free parameters the inclination $i$, the secondary star's temperature $T_2$, and the respective bandpass luminosities of the secondary star $L_2$. The W--D program iterated through the Levenberg--Marquardt procedure \citep{Levenberg1944, Marquardt1963} to find the best solution, as well as the mean residual, within a given number of iterations. In the second step, we repeated the same procedure for different $q$ values, from 0.05 to 10. The step width used was variable so as to balance the need for our computational resources and the resulting numerical precision (step widths of 0.02, 0.05, and 0.25 from $q=0.05$ to 0.5, from $q=0.5$ to 2, and from $q=2$ to 10, respectively). The total number of fixed $q$ values was 85. Note that, under certain conditions, the W--D code did not converge. We skipped the corresponding $q$ value and continued the calculation from the next $q$ value. CBs with fewer than 60 $q$ values were removed from our sample and subsequently ignored. Having thus obtained the best $q$ value, we relaxed the constraint on the mass ratio and carried out a final run based on all final, adjusted parameter values simultaneously to calculate the respective standard errors. 

\subsection{Absolute parameters}

Thanks to the high-precision CB PLRs derived by \citet{2018ApJ...859..140C}, we can now derive accurate absolute magnitudes for our sample CBs. To derive the absolute parameters, such as a system's semi-major axis ($A$) and its absolute stellar component masses, we must adopt a number of basic assumptions, imposed by the lack of spectroscopic data. We hence assumed that the luminosities and masses of the primary stars are commensurate with loci on the zero-age MS (ZAMS). This is a reasonable assumption \citep{2005ApJ...629.1055Y}. \citet{2013MNRAS.430.2029Y} compiled a list of 100 CBs with well-determined parameters and found that their primary components are more similar to normal MS stars than the systems' secondary components. These authors found that the primary stars occupy loci in both the $M$--$L$ and $M$--$R$ diagrams that make them resemble ZAMS stars as if they were detached eclipsing binaries, while the secondary stars do not exhibit such properties.

Using the luminosity fraction of the primary star $f_\mathrm{V}=L_\mathrm{V1} / (L_\mathrm{V1} + L_\mathrm{V2})$ derived in the previous section, for each CB we calculated the $V$-band luminosities of both component stars. Next, we converted these $V$-band luminosities to bolometric luminosities using the relevant bolometric correction \citep[BC;][]{2013ApJS..208....9P}. We obtained the $\mathrm{BC}_V$ for each star based on the effective temperature derived in Section~\ref{sec:qsearch}, i.e., $M_{i, \mathrm{bol}} = M_{i, V} + \mathrm{BC}_V(T_i)$, where $i$ corresponds to 1 or 2 in reference to the primary and secondary stars, respectively. We subsequently used the $M$--$L$ relation \citep[$L\propto M^{4.216}$;][]{2013MNRAS.430.2029Y} to infer the masses of the primary stars. The intercept of the $M$--$L$ relation was derived by fitting the CBs in the \citet{2013MNRAS.430.2029Y} catalog. The masses of the secondary stars were then determined based on the best-fitting mass ratios (Section~\ref{sec:qsearch}). The orbital major axes, $A$, were converted to absolute units using Kepler's Third Law. Therefore, we can deduce the absolute radii of the primary and secondary stars ($r_1$, $r_2$) based on their relative measurements ($r_1/A$, $r_2/A$). The errors associated with these absolute parameters were calculated through error propagation analysis.

\subsection{Selection Criteria \label{sec:selection}}

It is widely acknowledged that the $q$ values derived from spectroscopic studies may be different from those based on photometric analyses \citep[e.g.,][]{2005ApJ...629.1055Y}. \citet{2001AJ....122.1007R} have pointed out that the reliable method to determine the mass ratio should be based on radial velocity observations. In that case, the $q_\mathrm{sp}$ parameter is given by the ratio of velocity semi-amplitudes of both components. In fact, the $q_\mathrm{ph}$ parameter, i.e., the mass ratio obtained from light-curve analysis alone, might not be reliable. Spectroscopic $q_\mathrm{sq}$ values are usually preferred if the results are not mutually consistent.

However, determination of $q_\mathrm{ph}$ has been shown to be reliable nevertheless for the special conditions pertaining to systems exhibiting total eclipses \citep{1972MNRAS.156...51M, 1978ApJ...224..885W, 2001AJ....122.1007R}. In this case, the depth of the light-curve minima primarily depends on the mass ratio and much less on the fill-out factor. Combined with the duration of the totality, which allows for an estimation of the system's inclination, fully eclipsing binaries can break the degeneracy among the different physical parameters and yield an accurate mass ratio. In the ground-breaking study of \citet{2005Ap&SS.296..221T}, the authors simulated the light curves for various physical parameters and demonstrated that the eclipse properties (complete versus partial) govern photometric mass ratios for over-contact and semi-detached binaries. Only for CBs exhibiting total eclipses can accurate radii be derived based on Roche geometry, which hence results in accurate $q_\mathrm{ph}$ parameters. Subsequently, \citet{2013CoSka..43...27H} expanded the simulations to cover the full parameter space spanned by the mass ratio, the orbital inclination, and the fill-out factor to investigate the uniqueness of the photometric light-curve solutions. They addressed the importance of the presence of third light and also confirmed the result of \citet{2005Ap&SS.296..221T} that $q_\mathrm{ph}$ is robust for fully eclipsing over-contact and semi-detached systems. Under these circumstances, the severe degeneracy among multiple physical parameters, most notably between the mass ratio and the fill-out factor, can be broken.

Therefore, we applied additional selection criteria to our sample
  CBs to obtain a highly reliable sample. First, we visually checked
  the best-fitting solutions and excluded those that did not match
  well. The light curve of a typical CB should exhibit continuous
  brightness variations as a function of time and have nearly equal
  eclipse depths. In our next step, we neglected all CBs with
  inclinations below 70\degr. \citet{2013CoSka..43...27H} pointed out
  that the number of similar (i.e., degenerate) light curves decreases
  with increasing inclination, and so photometric light curves are not
  effective tools to analyze systems seen under low inclinations. CBs
  characterized by a large tilt of their orbital plane with respect to
  the observer ($i > 70\degr$) can have substantial variations in
  their brightness because of orbital eclipses. The final selection
  criterion was that only fully eclipsing systems were included in the
  final catalog to ensure a robust determination of the mass ratio,
  $q_\mathrm{ph}$. To achieve this, we regarded CBs with inclination
  angles $i > \arccos\left|(r_1-r_2)/A\right|$ to have total eclipses
  and their $q_\mathrm{ph}$ to be well-determined. A side effect of
  applying this criterion is that it will inevitably disfavor
  high-mass-ratio CBs. Therefore, a deficiency of CBs with $q\sim 1$
  was expected. Our final catalog includes 2335 CBs. The relative and
  absolute physical parameters derived are included in
  Tables~\ref{tab:relparam} and \ref{tab:absparam}, respectively.

\begin{deluxetable*}{lCCCCCCCCc}
\tablecaption{Relative Physical Parameters of our Sample CBs\label{tab:relparam}}
\tablewidth{0pt}
\tablehead{ \colhead{ID}&\colhead{Period}&\colhead{$T_1$\tablenotemark{a}}&\colhead{$T_2$\tablenotemark{a}} & \colhead{$i$} & \colhead{$\Omega$\tablenotemark{b}} & \colhead{$q$} & \colhead{$f$ \tablenotemark{c}} & \colhead{$L_\mathrm{V1} / L_\mathrm{V,tot}$} & \colhead{Subtype}\\ \colhead{} & \colhead{(day)} & \colhead{(K)} & \colhead{(K)} & \colhead{(\degr)} & \colhead{} & \colhead{} & \colhead{} & \colhead{} & \colhead{} }
\startdata 
CSS\_J223201.5+342945 & 0.27674 & 5330 \pm 38 & 5771 & 88.68 \pm 3.50 & 6.18 \pm 0.09 & 0.36 \pm 0.09 & 0.17 \pm 0.15 & 0.62 \pm 0.01 & W \\ 
CSS\_J090725.9-032447 & 0.36441 & 5845 \pm 32 & 6008 & 80.76 \pm 1.42 & 7.40 \pm 0.08 & 0.27 \pm 0.09 & 0.31 \pm 0.13 & 0.74 \pm 0.01 & W \\ 
CSS\_J223244.6+322638 & 0.29430 & 5086 \pm 35 & 5409 & 80.55 \pm 2.98 & 7.39 \pm 0.06 & 0.27 \pm 0.09 & 0.33 \pm 0.09 & 0.70 \pm 0.01 & W \\ 
CSS\_J165813.7+390911 & 0.27311 & 4806 \pm 28 & 5022 & 86.40 \pm 1.85 & 9.62 \pm 0.05 & 0.18 \pm 0.09 & 0.25 \pm 0.07 & 0.78 \pm 0.01 & W \\ 
CSS\_J001546.9+231523 & 0.27125 & 5614 \pm 37 & 6157 & 72.64 \pm 1.28 & 13.08 \pm 0.10 & 0.12 \pm 0.09 & 0.42 \pm 0.15 & 0.81 \pm 0.01 & W \\ 
CSS\_J222607.8+062107 & 0.39705 & 6229 \pm 128 & 6587 & 75.82 \pm 3.27 & 8.14 \pm 0.23 & 0.22 \pm 0.09 & 0.63 \pm 0.36 & 0.74 \pm 0.02 & W \\ 
CSS\_J042755.0+060421 & 0.30188 & 5380 & 5602 \pm 41 & 73.84 \pm 1.67 & 1.88 \pm 0.02 & 0.09 \pm 0.09 & 0.86 \pm 0.28 & 0.86 \pm 0.01 & W \\ 
CSS\_J080529.8+005305 & 0.35095 & 5492 \pm 50 & 5699 & 73.97 \pm 1.61 & 10.84 \pm 0.08 & 0.15 \pm 0.09 & 0.23 \pm 0.12 & 0.81 \pm 0.01 & W \\ 
CSS\_J225217.2+381800 & 0.34740 & 5635 \pm 30 & 5902 & 76.30 \pm 1.04 & 7.40 \pm 0.08 & 0.27 \pm 0.09 & 0.30 \pm 0.13 & 0.72 \pm 0.01 & W \\ 
CSS\_J163458.9-003336 & 0.30051 & 5380 \pm 38 & 5601 & 74.01 \pm 1.16 & 8.78 \pm 0.08 & 0.20 \pm 0.09 & 0.60 \pm 0.12 & 0.76 \pm 0.01 & W \\ 
CSS\_J012559.7+203404 & 0.39018 & 5567 \pm 35 & 5764 & 78.89 \pm 1.51 & 7.95 \pm 0.05 & 0.24 \pm 0.09 & 0.44 \pm 0.07 & 0.75 \pm 0.01 & W \\ 
CSS\_J041633.5+223927 & 0.31344 & 5469 & 5745 \pm 43 & 80.30 \pm 1.91 & 2.21 \pm 0.02 & 0.21 \pm 0.09 & 0.37 \pm 0.18 & 0.75 \pm 0.01 & W \\ 
CSS\_J145924.5-150145 & 0.45256 & 6164 & 5976 \pm 39 & 83.00 \pm 1.43 & 2.46 \pm 0.02 & 0.31 \pm 0.09 & 0.17 \pm 0.10 & 0.77 \pm 0.01 & A \\ 
CSS\_J051056.2+041919 & 0.38999 & 6627 & 6325 \pm 60 & 70.57 \pm 1.33 & 1.94 \pm 0.02 & 0.11 \pm 0.09 & 0.65 \pm 0.26 & 0.89 \pm 0.01 & A \\ 
CSS\_J130111.2-132012 & 0.36574 & 6061 & 6046 \pm 56 & 88.65 \pm 1.84 & 1.93 \pm 0.01 & 0.11 \pm 0.09 & 0.87 \pm 0.19 & 0.86 \pm 0.01 & A \\ 
CSS\_J130425.1-034619 & 0.23496 & 4703 & 4675 \pm 16 & 89.34 \pm 1.38 & 2.86 \pm 0.01 & 0.50 \pm 0.09 & 0.07 \pm 0.05 & 0.66 \pm 0.01 & A \\ 
CSS\_J141923.2-013522 & 0.31157 & 6701 & 6312 \pm 89 & 88.02 \pm 2.29 & 2.12 \pm 0.02 & 0.17 \pm 0.09 & 0.35 \pm 0.20 & 0.86 \pm 0.01 & A \\ 
CSS\_J065701.5+365255 & 0.30175 & 5226 & 5116 \pm 31 & 76.58 \pm 0.81 & 2.32 \pm 0.02 & 0.25 \pm 0.09 & 0.23 \pm 0.10 & 0.79 \pm 0.01 & A \\ 
CSS\_J162327.1+031900 & 0.47456 & 7187 & 5845 \pm 58 & 78.67 \pm 1.31 & 1.90 \pm 0.01 & 0.09 \pm 0.09 & 0.52 \pm 0.22 & 0.95 \pm 0.00 & B \\ 
CSS\_J153855.6+042903 & 0.36036 & 6788 & 5507 \pm 39 & 84.22 \pm 1.31 & 2.18 \pm 0.01 & 0.19 \pm 0.09 & 0.22 \pm 0.12 & 0.92 \pm 0.00 & B \\ 
\enddata
\tablenotetext{}{(This table is available in its entirety in machine-readable form in the online journal. A portion is shown here for guidance regarding its form and content.)}
\tablenotetext{a}{Temperatures without uncertainty estimates were derived using the photometric method described in Section~\ref{sec:temp}, while values with uncertainties were obtained from the W--D code.}
\tablenotetext{b}{$\Omega = \Omega_1 = \Omega_2$.}
\tablenotetext{c}{Fill-out factor, defined by \citet{1973AcA....23...79R}: $f = (\Omega - \Omega_\mathrm{o}) / (\Omega_\mathrm{i} - \Omega_\mathrm{i})$, where $\Omega_\mathrm{i}$ and $\Omega_\mathrm{o}$ are the inner and outer Lagrangian surface potential values, respectively.}
\end{deluxetable*}

\begin{deluxetable*}{lCCCCCCC}
\tablecaption{Absolute Physical Parameters of our Sample CBs\label{tab:absparam}}
\tablewidth{0pt}
\tablehead{ \colhead{ID}& \colhead{$m_1$}&\colhead{$m_2$} & \colhead{$r_1$} & \colhead{$r_2$} & \colhead{$L_1$} & \colhead{$L_2$} & \colhead{$A$}\\ \colhead{} & \colhead{($M_\odot$)} & \colhead{($M_\odot$)} & \colhead{($R_\odot$)} & \colhead{($R_\odot$)} & \colhead{($L_\odot$)} & \colhead{($L_\odot$)} & \colhead{($R_\odot$)} }
\startdata
CSS\_J223201.5+342945 & 1.04 \pm 0.03 & 0.38 \pm 0.09 & 0.96 \pm 0.03 & 0.61 \pm 0.01 & 0.74 \pm 0.17 & 0.41 \pm 0.10 & 2.01 \pm 0.04 \\ 
CSS\_J090725.9-032447 & 1.30 \pm 0.03 & 0.35 \pm 0.11 & 1.30 \pm 0.04 & 0.73 \pm 0.02 & 1.90 \pm 0.47 & 0.67 \pm 0.16 & 2.53 \pm 0.06 \\ 
CSS\_J223244.6+322638 & 1.06 \pm 0.03 & 0.28 \pm 0.09 & 1.06 \pm 0.03 & 0.60 \pm 0.01 & 0.82 \pm 0.19 & 0.33 \pm 0.08 & 2.06 \pm 0.05 \\ 
CSS\_J165813.7+390911 & 1.09 \pm 0.03 & 0.20 \pm 0.09 & 1.05 \pm 0.03 & 0.49 \pm 0.01 & 0.92 \pm 0.24 & 0.24 \pm 0.06 & 1.93 \pm 0.05 \\ 
CSS\_J001546.9+231523 & 1.09 \pm 0.03 & 0.13 \pm 0.09 & 1.09 \pm 0.03 & 0.44 \pm 0.01 & 0.92 \pm 0.23 & 0.20 \pm 0.05 & 1.89 \pm 0.05 \\ 
CSS\_J222607.8+062107 & 1.49 \pm 0.03 & 0.33 \pm 0.13 & 1.51 \pm 0.05 & 0.81 \pm 0.02 & 3.35 \pm 0.65 & 1.15 \pm 0.24 & 2.77 \pm 0.07 \\ 
CSS\_J042755.0+060421 & 1.21 \pm 0.04 & 0.11 \pm 0.10 & 1.27 \pm 0.04 & 0.47 \pm 0.01 & 1.39 \pm 0.43 & 0.22 \pm 0.07 & 2.07 \pm 0.06 \\ 
CSS\_J080529.8+005305 & 1.28 \pm 0.03 & 0.20 \pm 0.11 & 1.32 \pm 0.04 & 0.58 \pm 0.01 & 1.80 \pm 0.44 & 0.40 \pm 0.10 & 2.39 \pm 0.06 \\ 
CSS\_J225217.2+381800 & 1.31 \pm 0.03 & 0.35 \pm 0.11 & 1.27 \pm 0.04 & 0.71 \pm 0.02 & 2.00 \pm 0.47 & 0.74 \pm 0.18 & 2.47 \pm 0.06 \\ 
CSS\_J163458.9-003336 & 1.17 \pm 0.03 & 0.23 \pm 0.10 & 1.16 \pm 0.04 & 0.60 \pm 0.01 & 1.24 \pm 0.33 & 0.35 \pm 0.09 & 2.12 \pm 0.05 \\ 
CSS\_J012559.7+203404 & 1.38 \pm 0.03 & 0.32 \pm 0.12 & 1.42 \pm 0.04 & 0.77 \pm 0.02 & 2.46 \pm 0.56 & 0.81 \pm 0.18 & 2.69 \pm 0.06 \\ 
CSS\_J041633.5+223927 & 1.34 \pm 0.04 & 0.28 \pm 0.11 & 1.22 \pm 0.04 & 0.63 \pm 0.02 & 2.20 \pm 0.68 & 0.69 \pm 0.21 & 2.28 \pm 0.06 \\ 
CSS\_J145924.5-150145 & 1.55 \pm 0.04 & 0.48 \pm 0.13 & 1.55 \pm 0.05 & 0.92 \pm 0.02 & 3.97 \pm 0.99 & 1.22 \pm 0.30 & 3.14 \pm 0.07 \\ 
CSS\_J051056.2+041919 & 1.48 \pm 0.04 & 0.16 \pm 0.13 & 1.57 \pm 0.05 & 0.62 \pm 0.02 & 3.26 \pm 0.78 & 0.42 \pm 0.11 & 2.65 \pm 0.07 \\ 
CSS\_J130111.2-132012 & 1.38 \pm 0.03 & 0.15 \pm 0.12 & 1.49 \pm 0.05 & 0.61 \pm 0.02 & 2.49 \pm 0.57 & 0.40 \pm 0.09 & 2.48 \pm 0.07 \\ 
CSS\_J130425.1-034619 & 0.95 \pm 0.03 & 0.47 \pm 0.08 & 0.80 \pm 0.02 & 0.59 \pm 0.01 & 0.50 \pm 0.13 & 0.26 \pm 0.07 & 1.80 \pm 0.04 \\ 
CSS\_J141923.2-013522 & 1.31 \pm 0.05 & 0.22 \pm 0.11 & 1.23 \pm 0.04 & 0.57 \pm 0.02 & 1.97 \pm 0.68 & 0.33 \pm 0.11 & 2.23 \pm 0.06 \\ 
CSS\_J065701.5+365255 & 1.11 \pm 0.03 & 0.28 \pm 0.09 & 1.09 \pm 0.03 & 0.59 \pm 0.01 & 0.96 \pm 0.21 & 0.26 \pm 0.06 & 2.11 \pm 0.05 \\ 
CSS\_J162327.1+031900 & 1.74 \pm 0.04 & 0.16 \pm 0.15 & 1.91 \pm 0.06 & 0.69 \pm 0.02 & 6.48 \pm 1.46 & 0.36 \pm 0.09 & 3.17 \pm 0.09 \\ 
CSS\_J153855.6+042903 & 1.44 \pm 0.05 & 0.27 \pm 0.12 & 1.37 \pm 0.04 & 0.66 \pm 0.02 & 2.94 \pm 0.96 & 0.30 \pm 0.10 & 2.55 \pm 0.07 \\
\enddata
\tablenotetext{}{(This table is available in its entirety in machine-readable form in the online journal. A portion is shown here for guidance regarding its form and content.)}
\end{deluxetable*} 

\section{Validation \label{sec:validation}}

The reliability of our results is predominantly determined by the
  quality of our measurements, which renders validation of great
  importance. To assess the performance quality of our method, two
  tests were designed, to evaluate the final accuracy and precision,
  respectively. `Accuracy' here refers to how close our derived values
  are to the `true' value, while `precision' reflects how close our
  results are to each other. Sections~\ref{sec:accuracy} and
  \ref{sec:precision} address, respectively, the accuracy and
  precision of the physical parameters $q$.

In our accuracy test, we compared our results with spectroscopic
  measurements from the literature. In general, $q_\mathrm{sp}$ values
  based on spectroscopic velocity curves are usually considered the
  `correct' means to evaluate the `true' mass ratios, while
  $q_\mathrm{ph}$ might be influenced by other properties. Thus, such
  a direct comparison can tell us directly whether there are any
  discrepancies between our results and the `true' values, and obtain
  a reasonable approximation to the uncertainties associated with a
  range of physical parameters. In the precision test, we applied our methodology to ASAS-SN
  data to check whether the parameters derived from various sky
  surveys are biased with respect to each other. This way, we can
  assess the coherence of our measurements across different data
  sources.

\subsection{Accuracy testing with spectroscopic measurements\label{sec:accuracy}}

In this section, we will perform a direct comparison between our results and literature data. Since our CSS-based CBs are generally fainter than the CBs in the \citet{2003CoSka..33...38P} catalog, we did not find any matching candidates. Instead, we collected ASAS-SN CB light curves for which literature measurements from \citet{2003CoSka..33...38P} were available. These CBs were cross-matched with APASS
  and \textit{Gaia} DR2 (based on their coordinates) to derive color
  indices and absolute distances. Next, we derived the light curve
  solution and selected a sample with reliable measurements adopting
  the same selection criteria as before. The final step was to
  estimate the scatter in various parameters (e.g., $q$ and $f$)
  compared with their values in the literature. The systematic
  uncertainty estimated from the ASAS-SN data also applies to our
  CSS-based results, because both surveys share the same passband
  ($V$), while the typical sampling cadence and the photometric
  uncertainties are comparable.

\begin{figure*}[ht]
\plottwo{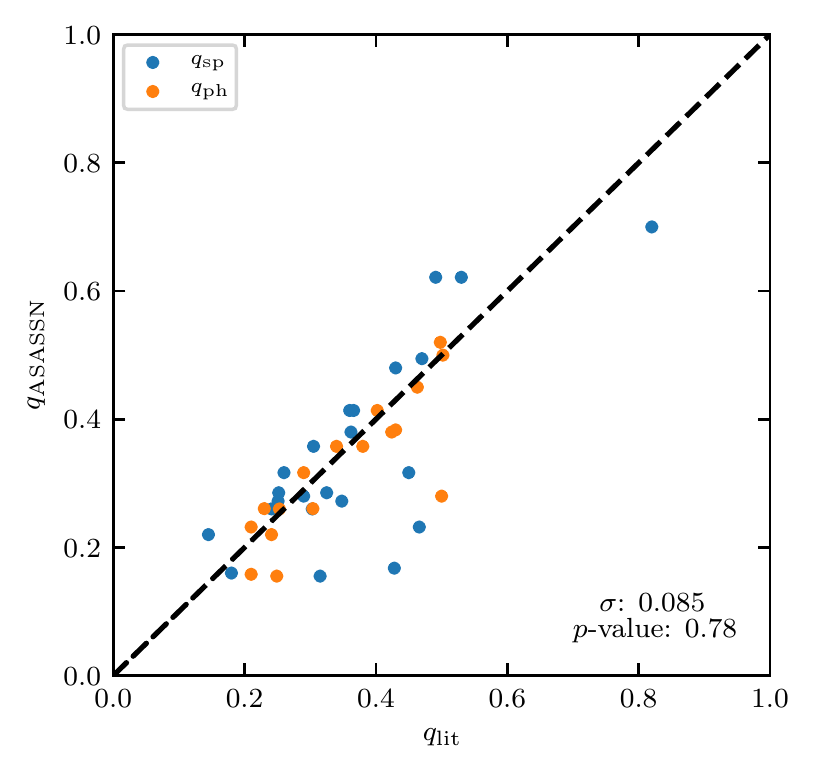}{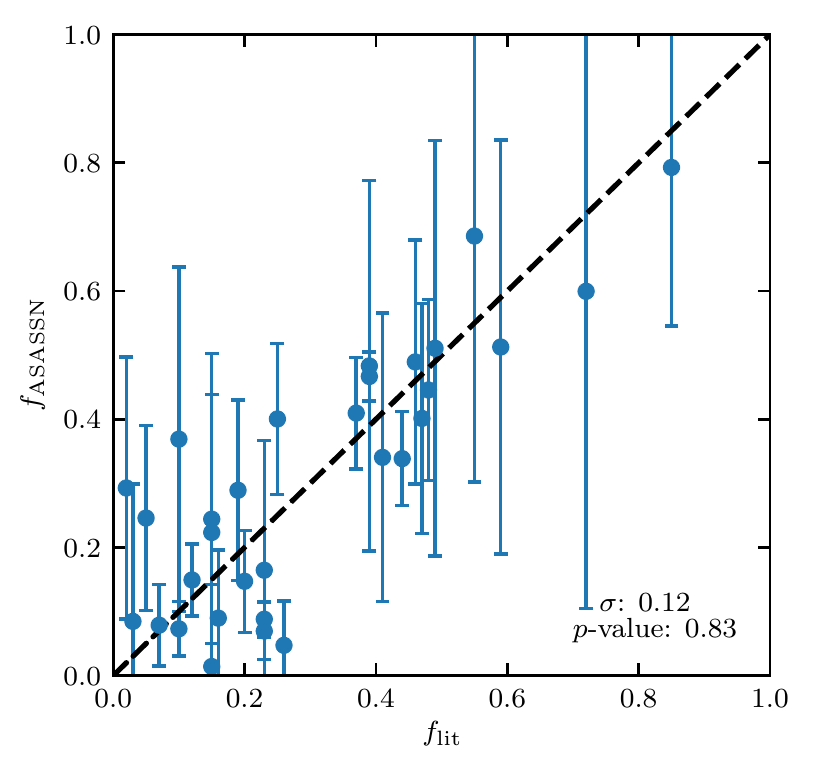}
\caption{Comparisons of (left) mass ratio $q$ and (right) fill-out
  factor $f$ (right) of literature values and the solutions we derived
  from ASAN-SN data. The black dashed lines are the one-to-one linear
  relations for $q$ and $f$. The root-mean-square error, $\sigma$, and
  the Pearson correlation coefficients for these parameters are
  included in the bottom right-hand corners of the panels. In the
  left-hand panel, mass ratios from spectroscopic and photometric
  sources are marked in blue and orange colors,
  respectively. \label{fig:comp}}
\end{figure*}

In the left-hand panel of Fig.~\ref{fig:comp}, we present a
  comparison of the mass ratios $q$ of literature values and the
  solutions we derived from ASAN-SN. Note that we also included
  literature results based on photometric light curves. This is a
  reasonable practice, since \citet[][their
    Fig.~1]{2003CoSka..33...38P} confirmed the consistency of
  $q_\mathrm{ph}$ and $q_\mathrm{sp}$ for total eclipses. Forty of the
  CBs we obtained light curve solutions for based on ASAS-SN data had
  either $q_\mathrm{sp}$ or $q_\mathrm{ph}$ measurements
  available. The mass ratios calculated based on ASAS-SN light curves,
  $q_\mathrm{ASASSN}$, are in good agreement with their literature
  counterparts, $q_\mathrm{lit}$. The corresponding Pearson
  correlation coefficient is 0.78, indicating a strong linear
  correlation between both measurements. The mean difference in the
  mass ratios, $\Delta q = q_\mathrm{ASASSN}-q_\mathrm{lit} = -0.02$,
  which is only a fraction of the r.m.s. error ($\sigma =
  0.085$). This good agreement implies that our measurements of the
  mass ratios are fully consistent with the `true' values and there
  are no significant discrepancies. Therefore, we adopted the scatter,
  $\sigma$, as the actual uncertainty in the mass ratio for our CSS
  data set.

We additionally checked our determinations of the fill-out
  factor, $f$, which may also suffer from degeneracies: see the
  right-hand panel of Fig.~\ref{fig:comp}. Except for some points with
  relatively large error bars, there is a good linear correlation
  between $f_\mathrm{lit}$ and $f_\mathrm{ASASSN}$. This is strong
  evidence supporting, based on the photometric precision of ASAS-SN
  (or CSS), that we can derive accurate measurements of physical
  parameters that are not severely biased.

\subsection{Precision testing with ASAS-SN\label{sec:precision}}

\begin{figure*}[ht]
\plotone{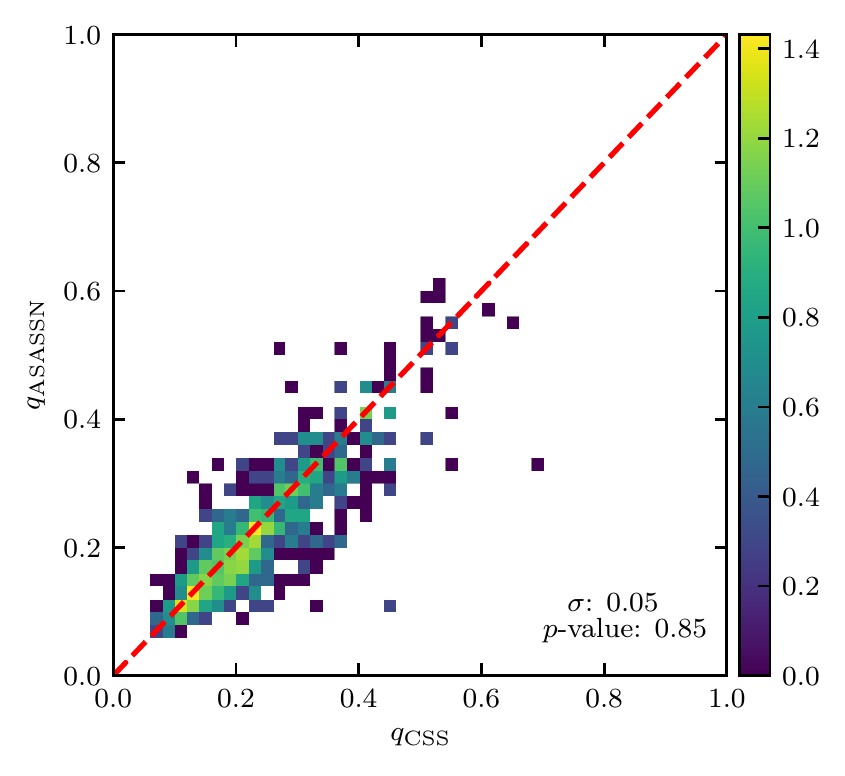}
\caption{Hess diagram of the mass ratios derived from CSS
  ($q_\mathrm{CSS}$) and ASAS-SN data ($q_\mathrm{ASASSN}$). Colors
  represent the logarithm of the number of objects in each bin. The
  red dashed line is the one-to-one linear relation. The
  root-mean-square error, $\sigma$, and Pearson correlation
  coefficients are included in the bottom right-hand cornerc of the
  panels.\label{fig:consist}}
\end{figure*}

We also performed a consistency test to verify whether our
  measurements are coherent among different surveys. A subsample of
  877 CBs was randomly selected from our catalog and we made a
  comparison of the physical parameters ($q$) derived based on CSS and
  those based on ASAS-SN data. The result of the comparison
  (Fig.~\ref{fig:consist}) is shown as a Hess diagram to better
  illustrate the relative density of data points. The mass ratio
  measurements demonstrate a remarkable consistency among various
  surveys. The scatter in this correlation ($\sigma = 0.05$) could be
  taken as the the internal error associated with our method, which is
  smaller than the $\sigma=0.08$ reported in
  Section~\ref{sec:accuracy}. This behavior is what one can expect
  when comparing with an external catalog. Although there is a lack of
  CBs with a high mass ratios, this test is sufficient to illustrate
  the coherence of our measurements, i.e., that it is not strongly
  biased by the photometric uncertainties. A more robust test could be
  done by comparison with a high-precision survey (e.g., the Zwicky
  Transient Factory, ZTF). However, the number of available objects
  with high-cadence light curves covering the entire phase space is
  limited. Therefore, we did not include a comparison with the ZTF,
  but we will explore the ZTF in a future paper.

\section{Discussion \label{sec:discussion}}

\subsection{CB subtypes}

Equipped with this information about the relative parameters of our
sample CBs, we now can classify them into several
subtypes. Traditionally, CBs are divided into two subtypes: A-type
systems (where the more massive star is hotter) and W-type systems
(where the less massive star is hotter). A further subdivision,
referred to as B-type CBs, has been proposed to describe systems that
exhibit a significant temperature difference between the primary and
secondary components \citep{1979ApJ...231..502L}. These latter CB
systems are in marginal contact with each other and cannot attain
thermal equilibrium. We adopted the criterion that B-type CBs should
exhibit a temperature difference between their components over
\unit[1000]{K}, while A- and W-type CBs are classified based on their
masses and temperatures. Our sample contains 1530 W-, 710
A-, and 95 B-type CBs.

\begin{figure*}[ht!]
\plotone{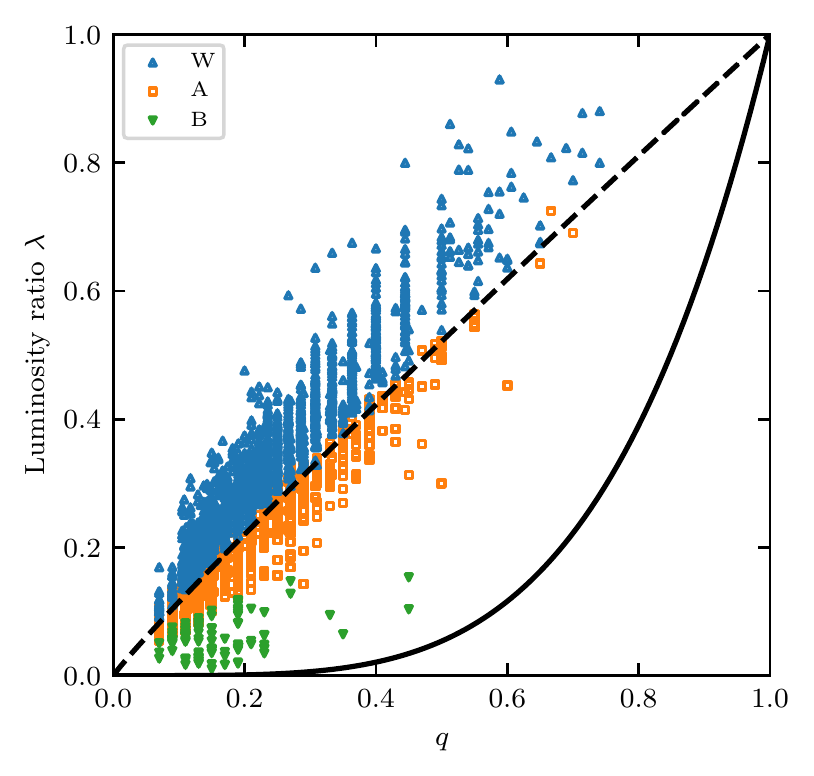}
\caption{Luminosity ratio ($\lambda$) versus mass ratio ($q$) distribution of our CB sample. Open orange squares, open blue triangles, and solid green triangles represent A-, W-, and B-type CBs, respectively. The solid line is the MS $M$--$L$, i.e., $\lambda = q^{4.216}$; the dashed line is Lucy's relation, $\lambda = q^{0.92}$. \label{fig:Lratio_q}}
\end{figure*}

Figure~\ref{fig:Lratio_q} shows the distribution of the bolometric luminosity ratio, $\lambda = L_2/L_1$, of our CBs as a function of $q$. The CB subtypes occupy different regions in the diagram. A- and W-type systems reside close to the correlation found by \citet{1968ApJ...153..877L}, $\lambda=q^{0.92}$. \citet{1968ApJ...153..877L} argued that the apparent ratio of the CBs' luminosities does not follow the MS relation, $\lambda=q^{4.216}$, but that it is instead proportional to the ratio of the surface areas. This suggests that mass exchange may be significant among A- and W-type systems. However, B-type CBs are located between Lucy's relation and the $\lambda=q^{4.216}$ line, in essence since B-type CBs are binary systems that have not yet attained thermal equilibrium. Note that this is different from our assumption for the primary stars adopted in the previous section. Here, we consider the luminosity ratios of the primary and secondary components. On the one hand, if the prevailing energy transfer is sufficient, they should have attained the same temperature but different sizes. On the other hand, if the energy transfer is not sufficient, both components resemble independently evolved stars, which would thus follow the $\lambda=q^{4.216}$ relation. In other words, we only ascertain whether the luminosity ratios follow either of the known trends. We also found that W-type CBs have generally higher luminosity ratios than their A-type counterparts for a given mass ratio. This is expected because the $T_2/T_1$ temperature ratio is higher for W-type systems.

\citet{2004A&A...426.1001C} introduced the concept of H-type CBs, characterized by high mass ratios, $q \ge 0.72$, which were found to exhibit different energy-transfer behaviors:
\begin{equation}
\beta = \frac{L_{1, \mathrm{obs}}}{L_{1, \mathrm{ZAMS}}},
\end{equation}
where $L_{1, \mathrm{obs}}$ is the observed luminosity of the primary star, $L_{1, \mathrm{obs}} = L_\mathrm{tot}/(1+q^{0.92}\left(\frac{T_2}{T_1}\right)^4) = L_\mathrm{tot}/(1+\lambda)$, following the model of \citet{1968ApJ...153..877L}, and $L_{1, \mathrm{ZAMS}}$ is the luminosity of the primary star if both stars follow the MS $M$--$L$ relation, $L_{1, \mathrm{ZAMS}} = L_\mathrm{tot}/(1+q^{4.216})$. It is straightforward to show that
\begin{equation}
\beta = \frac{1+q^{4.216}}{1+q^{0.92}\left(\frac{T_2}{T_1}\right)^4}=\frac{1 + \alpha \lambda^{4.58}}{1+\lambda}, \label{eq:transfer}
\end{equation}
where $\alpha=\left(\frac{T_1}{T_2}\right)^{18.3}$. Note that the $M$--$L$ relation we have adopted \citep{2013MNRAS.430.2029Y} is slightly different from that of \citet{2004A&A...426.1001C}, and hence the indices are not exactly the same. We adopted the former relation since it provided better fits to our data.

\begin{figure*}[ht!]
\includegraphics[width=0.33\textwidth]{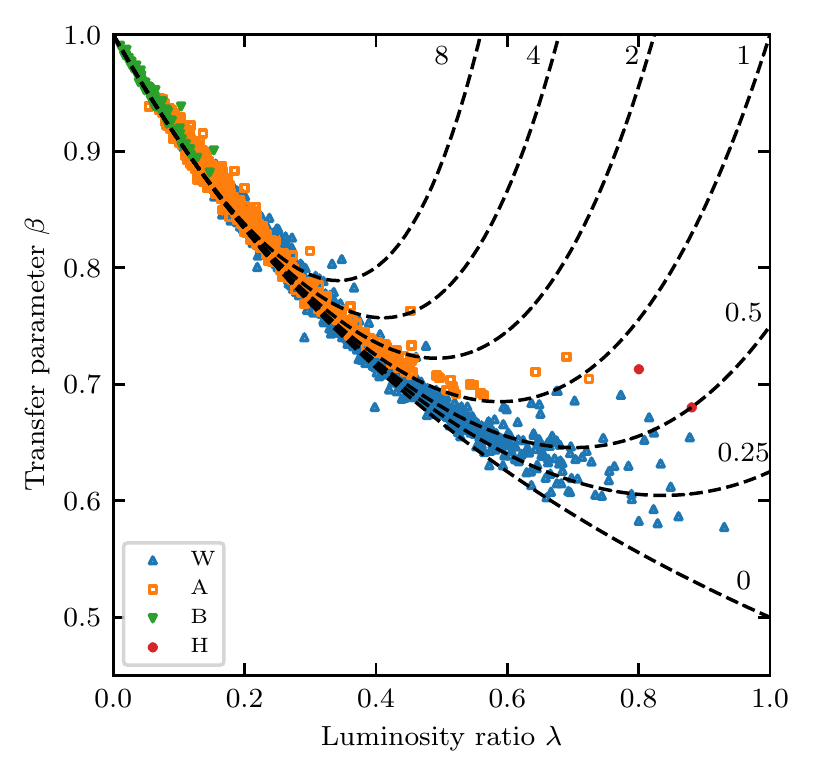}
\includegraphics[width=0.66\textwidth]{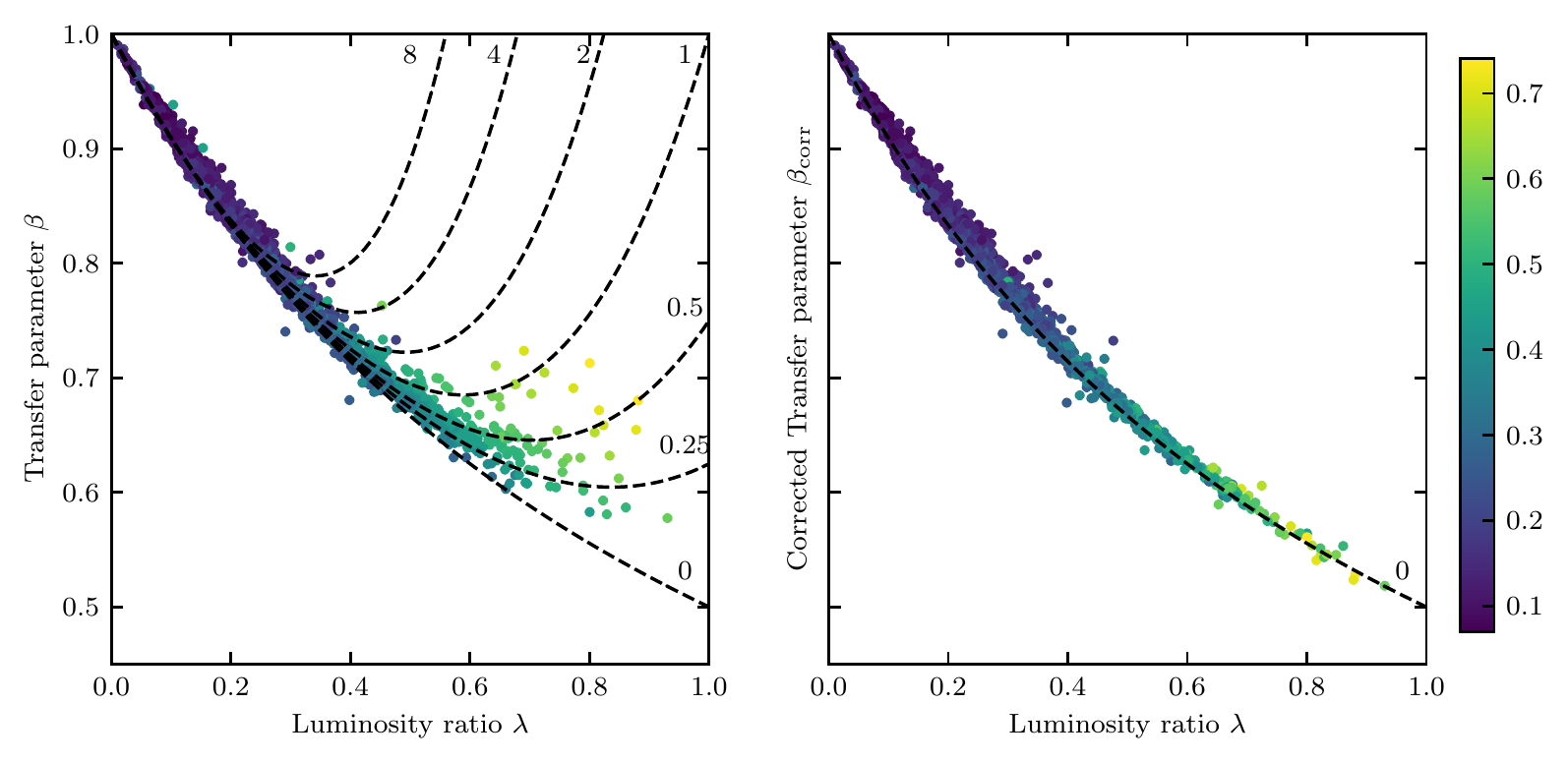}
\caption{(left) Transfer parameter $\beta$ versus luminosity ratio $\lambda$. Symbols are as in Fig.~\ref{fig:Lratio_q}. Dashed lines show the expected $\beta$ curves for different $\alpha$ values. Red solid dots represent H-type CBs, defined as high mass-ratio binary systems, $q > 0.72$. H-type CBs are clearly located away from the envelope ($\alpha = 0$). (middle) As the left-hand panel, but color-coded by mass ratio, $q$. (right) Corrected transfer parameter, $\beta_\mathrm{corr}$, versus $\lambda$. The color bar on the right applies to the middle and right-hand panels.\label{fig:transfer}}
\end{figure*}

We present the distribution of our CB sample's transfer parameters
$\beta$ versus their luminosity ratios $\lambda$ in the left-hand
panel of Fig.~\ref{fig:transfer}. We classified all systems with high
mass ratios ($q \ge 0.72$) as H-type stars. As expected, only two
  CBs were marked as H types due to our selection criterion aimed at
  only selecting CBs with total eclipses. For most systems, both
parameters exhibit a good correlation that can be represented well by
Eq.~\ref{eq:transfer}, with $\alpha$ ranging from 0.5 to 2. Note that
$\alpha$ depends sensitively on the ratio of the components' surface
temperatures, suggesting that the surface temperatures of the primary
and secondary stars in the majority of A-, B-, and W-type CBs are very
similar. These subtypes are enclosed by an envelope corresponding to
the minimum rate of transfer at a given luminosity ratio ($\alpha =
0$). It has been suggested \citep{2001OAP....14...33K} that the energy
transfer rate is a function of the luminosity of the secondary
star. However, \citet{2004A&A...426.1001C} found that the former
parameter is also related to the mass ratio. In the middle panel, we
redrew the figure by color-coding the data according to the CBs' mass
ratios. The deviation of high-$q$ CBs from the envelope ($\alpha
= 0$) shows a clearly increasing trend as $q$ becomes larger,
attaining significance for $q > 0.6$. In fact,
\citet{2004A&A...426.1001C} corrected their $\beta$ values to account
for the influence of different mass ratios, i.e., $\beta_\mathrm{corr}
= \beta - 0.54 q ^{4.1}$, leading to a correlation between
$\beta_\mathrm{corr}$ and the bolometric luminosity ratio. In the
right-hand panel, we adopted this practice and indeed confirmed their
results. That is, we did not find any evidence indicating that CBs
with mass ratios greater than 0.72 are special. Therefore, we did not
include H-type CBs as a subtype in our classification.

\subsection{Periods and evolutionary state}

One of the key parameters defining a given CB system is its orbital period, which is commonly used as a proxy for its evolutionary state \citep[e.g.,][]{2001MNRAS.328..635Q}. As mass transfer proceeds, a binary system's orbital separation continues to shrink, thus leading to a decrease in the orbital period.

\begin{figure}[ht!]
\plotone{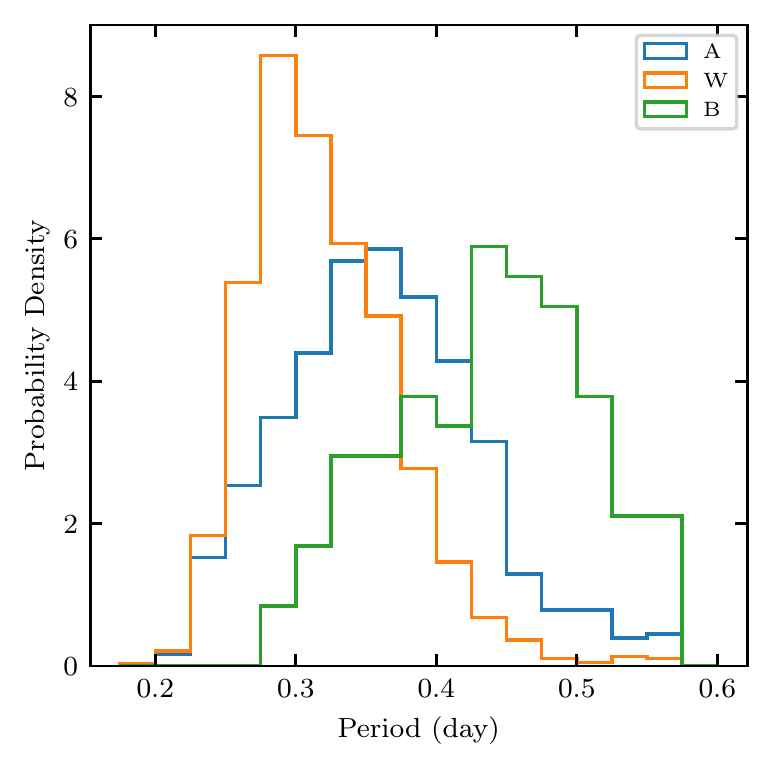}
\caption{Orbital-period distributions for different CB subtypes.\label{fig:period}}
\end{figure}

Figure~\ref{fig:period} shows the distribution of orbital periods for
the different CB subtypes. The period distribution of B-type CBs peaks
around \unit[0.45]{day}, which is distinct from the distributions of
the other subtypes. B-type CBs are likely in the non-thermal-contact
state of the relaxation oscillations and a semidetached phase
\citep{1979ApJ...231..502L}. Approximately one-quarter of B-type
systems have relatively short periods. However, note that the
  prevailing selection effects are rather complicated. In fact,
  they may favor the detection of systems exhibiting large
  amplitudes. On the other hand, high-$q$ CBs are likely rejected
  because of our focus on selecting objects exhibiting total
  eclipses.

Whether or not the CB subtypes represent an evolutionary sequence is
the subject of debate \citep{1996A&A...311..523M, 2005JKAS...38...43A,
  2006MNRAS.373.1483E, 2006MNRAS.372L..83G,
  2013MNRAS.430.2029Y}. Tentative evidence suggests that, if an
evolutionary sequence exists, it should reflect an evolution from A-
to W-type systems. \citet{2006MNRAS.372L..83G} found that A-type CBs
generally have longer periods compared with W-type systems for a given
orbital angular momentum. This supports the argument that evolution
from A- to W-type systems may be associated with simultaneous mass and
angular momentum loss. Evolution in the opposite direction is less
likely since there is no injection of mass or angular momentum from
outside of the CB systems. Figure~\ref{fig:period} shows that,
although the period distributions of A- and W-type CBs largely
overlap, A-type systems tend to have longer periods. Even though this
distribution has not been corrected for selection effects, there is no
evidence that A-type CBs are more affected by selection biases and,
therefore, this may reveal a general property of the period
distribution.

Our result supports the notion that A-type systems are less evolved than W-type systems, which might be because A-type CBs have not gone through the mass-reversal stage. However, a number of studies disagree with this scenario. \citet{1988MNRAS.231..341H} claimed that W-type CBs are not evolved MS stars and that A-type systems have almost reached the terminal MS age. \citet{2013MNRAS.430.2029Y} estimated that the initial masses of A- and W-type CBs are different by assuming that mass transfer starts near the terminal MS age. They found that semi-detached systems with a massive secondary component ($>\unit[1.8]{M_\odot}$) will form A-type CBs, while systems with a less massive secondary component ($<\unit[1.5]{M_\odot}$) will evolve to the contact phase because of the rapid evolution of angular momentum, and hence form W-type CBs. Thus, evolutionary connections among the various CB subtypes, if any, are still unclear.

\begin{figure}[ht!]
\includegraphics[width=0.33\textwidth]{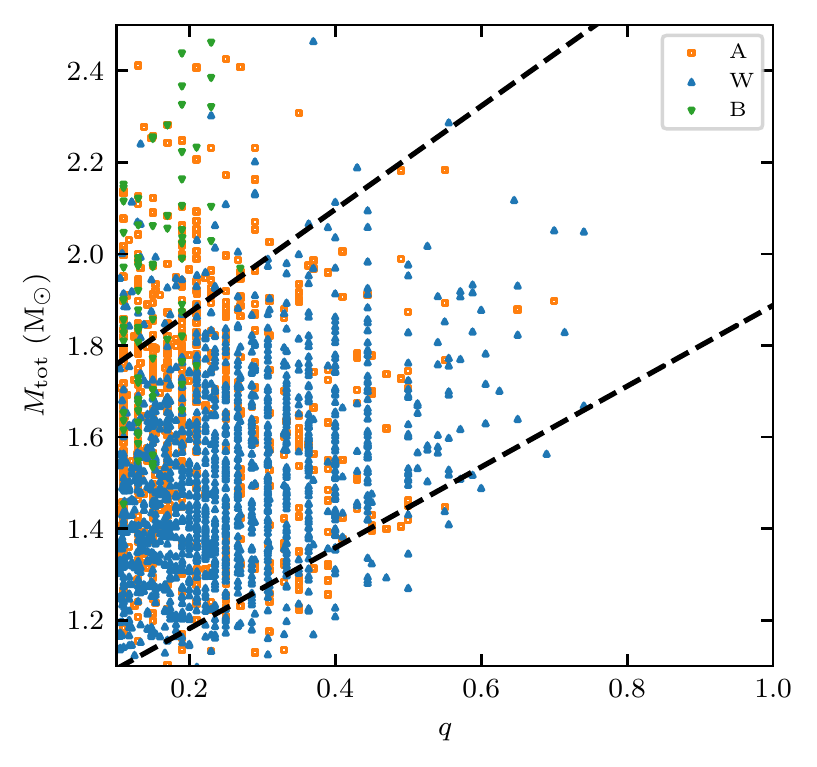}
\includegraphics[width=0.66\textwidth]{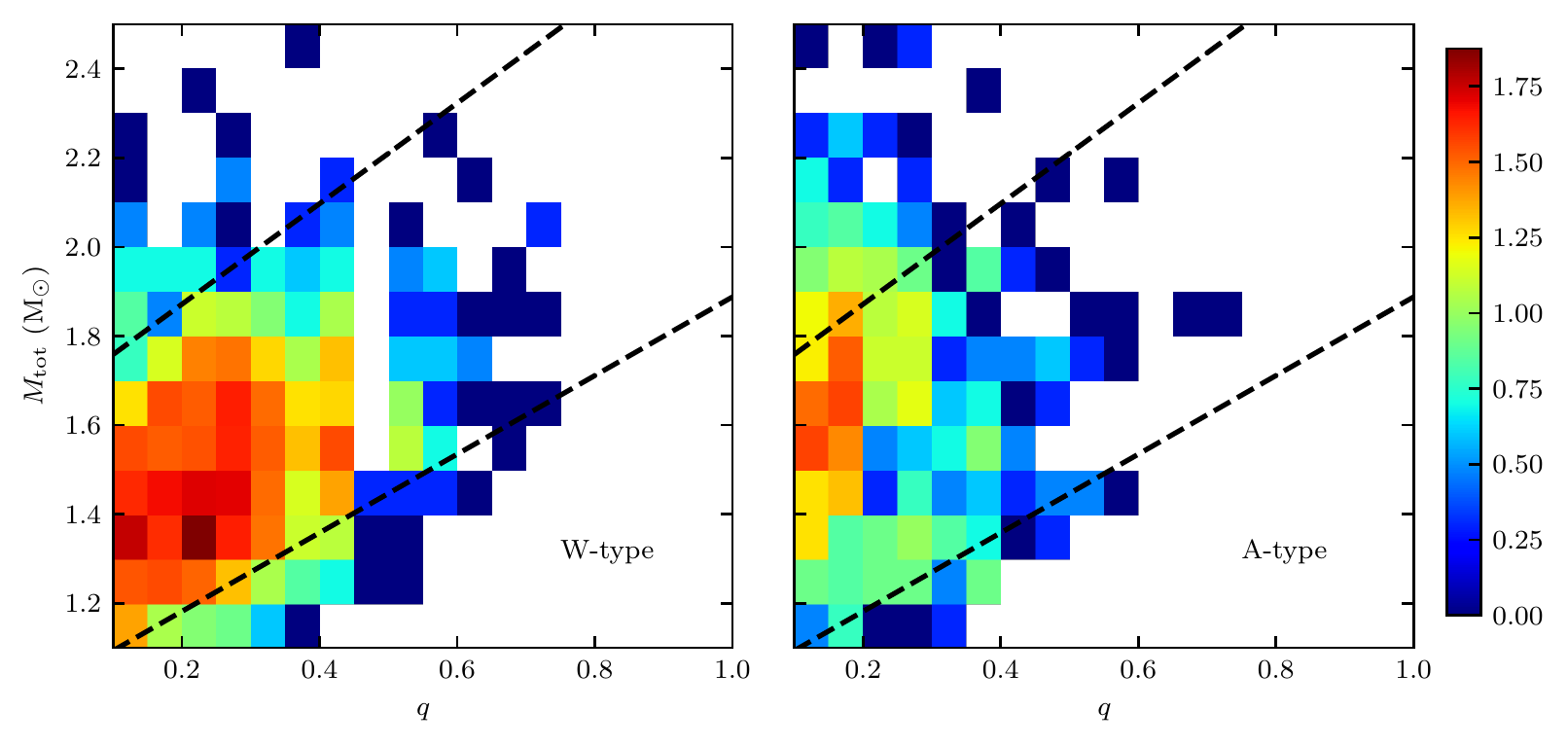}
\caption{Total CB mass, $M_\mathrm{tot}$, as a function of mass ratio,
  $q$. Symbols are as in Fig.~\ref{fig:Lratio_q}. Black dashed lines
  represent the edges of the strip defined by the 5\% and 95\%
  percentiles of $M_\mathrm{tot}$ for each $q$ bin, where
  $0.1\leqslant q \leqslant 1$ for bin steps of 0.05. The
    corresponding Hess diagrams for W- and A-type CBs are presented in
    the middle and right-hand panels, respectively. Colors represent
    the logarithm of the number of objects in each
    bin. \label{fig:q_m}}
\end{figure}

In Fig.~\ref{fig:q_m}, the total CB mass, $M_\mathrm{tot}$, is shown
as a function of $q$. Most A- and W-type CBs are located in a
strip. This region is delineated by the black dashed lines, defined by
the 5\% and 95\% percentiles of $M_\mathrm{tot}$ for each $q$ bin,
where $0.1\leqslant q \leqslant 1$ for bin steps of 0.05. This
feature, which was discovered by \citet{1996ASPC...90..280V}, has
subsequently been confirmed based on a sample of 130 CBs with
well-determined physical parameters
\citep{2008MNRAS.387...97L}. Moreover, the strip's lower boundary is
better defined than its upper boundary. In addition, a significant
fraction of B-type CBs lies beyond the strip, suggesting a rather
different evolutionary path for B subtypes. The slope of the best
linear fit to our sample CBs (excluding the B subtypes) is $\dd m/\dd
q = 0.57\pm 0.02$, which is consistent with
\citet{1996ASPC...90..280V} to within $\unit[1]{\sigma}$. However, close inspection revealed that the story might not be that
  simple. In the middle and right-hand panels of Fig.~\ref{fig:q_m},
  we show the Hess diagrams for W- and A-type CBs, overplotted with
  the same black dashed lines as in the left-hand panel. It is clear
  that although W-type CBs are located in a well-defined strip, a
  non-trivial fraction of A-type stars lie outside of this
  region. Moreover, the strip-like morphology for A-type CBs is much
  less obvious compared with their W-type counterparts, and even if
  similar boundaries exist for A-type CBs, the dominant slope appears
  different. This difference is also tentatively visible in
  \citet[][their Fig.~4]{2008MNRAS.387...97L}, where low-mass-ratio
  ($q \lessapprox0.6$) A-type CBs were generally found close to the
  high-mass boundary.

This trend, suggesting that (at least for W-type CBs) the lower
the total mass of the CBs is, the smaller their mass ratio becomes,
could be a natural product of their dynamical evolution in the absence
of mass reversal \citep{1976Ap&SS..39..447L, 1976ApJ...205..217F,
  1977MNRAS.179..359R, 1982A&A...109...17V}. However, other models
\citep{2006AcA....56..199S, 2007MNRAS.378..961P} imply that mass-ratio
reversal of the progenitors occurs during the system's evolution. Our
current sample may not allow us to differentiate between both
scenarios.

\citet{2008MNRAS.387...97L} claimed that W-type systems are generally
found in a region with intermediate-mass ratios between 0.3 and 0.7,
while A-type systems occur much less commonly in this area. Instead,
the latter are located in two separate regions of parameter space
($q\leqslant 0.5$ and $q\geqslant 0.7$). However, we do not see this
pattern in our sample, and we attribute the
\citet{2008MNRAS.387...97L} result to selection effects. Compared with
previous studies \citep[e.g.,][]{2004A&A...426.1001C}, our sample
contains a larger fraction of low mass-ratio CBs. Thus, there may be
some systematic differences between the samples; the effects of
  selection criteria have been discussed. Previous CB analyses were
usually based on small sample sizes ($\sim 100$ objects) and limited
to the solar neighborhood ($\lesssim\unit[300]{pc}$), while here our
sample is drawn from a larger volume, extending to distances of
$\unit[2-3]{kpc}$). We confirmed that if we limit our sample to the
solar neighborhood, the resulting mass-ratio distribution in the
  low-$q$ regime is similar to those published previously.

\subsection{Period--luminosity relations}

\begin{figure*}[ht!]
\plotone{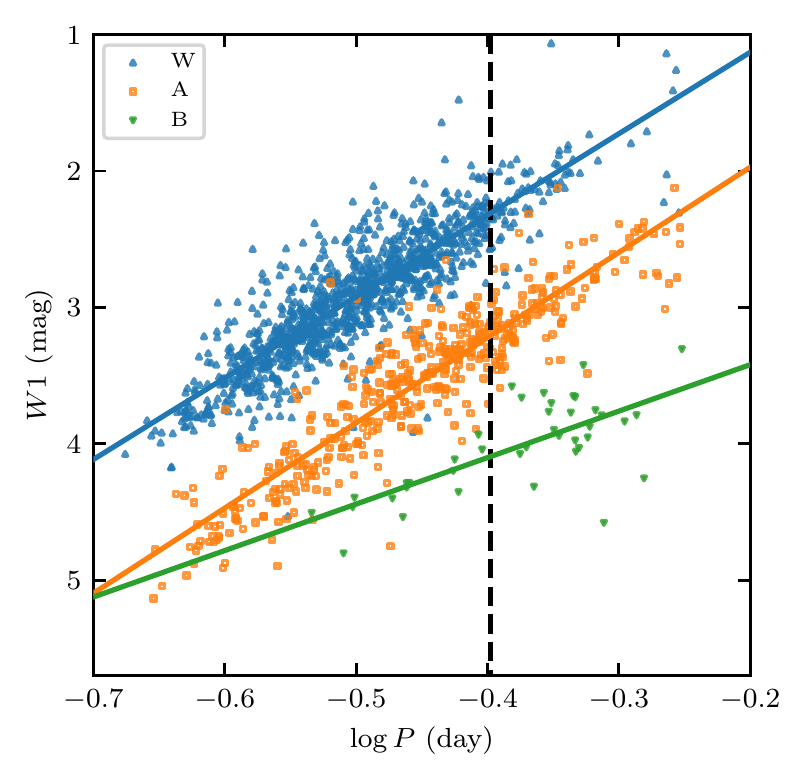}
\caption{PLRs based on distances from \textit{Gaia} DR 2 for different CB subtypes. A- and B-type CBs are offset by \unit[1]{mag} and \unit[2]{mag}, respectively. The dashed line represents $P = \unit[0.4]{day}$, for which the intercepts have been calculated. \label{fig:plr}}
\end{figure*}

Equipped with such a large CB sample, we can now study whether there are any systematic differences in the PLRs for different subtypes. Using the distances estimated by \citet{2018AJ....156...58B} based on \textit{Gaia} DR 2 \citep{2018A&A...616A...1G} parallax measurements, we constructed the PLRs for W-, A-, and B-type CBs: see Fig.~\ref{fig:plr}. \citet{2018ApJ...859..140C} found that $W1$-band distances are better than their $G$-band counterparts because the mid-infrared $W1$ band is less affected by extinction and metallicity variations.

The corresponding best-fitting PLRs are:
\begin{align*}
\mathrm{W}: M_{W1,\mathrm{mean}} &= (-5.97\pm 0.10) (\log P - \log 0.4) + (2.31 \pm 0.01),\qquad \sigma = \unit[0.23]{mag} \qquad N = 1130\\
\mathrm{A}: M_{W1,\mathrm{mean}} &= (-6.25\pm 0.13) (\log P - \log 0.4) + (2.21 \pm 0.01),\qquad \sigma = \unit[0.24]{mag} \qquad N = 457\\
\mathrm{B}: M_{W1,\mathrm{mean}} &= (-3.41\pm 0.56) (\log P - \log 0.4) + (2.09 \pm 0.04),\qquad \sigma = \unit[0.24]{mag} \qquad N = 40
\end{align*}
These PLR slopes for A- and W-type CBs are consistent with the slopes derived by \citet{2018ApJ...859..140C}, to within $\unit[2]{\sigma}$. This shows that CBs obey rather tight correlations between their periods and luminosities. The reason that the scatter ($\sigma$) resulting from our fits is larger than that derived by \citet{2018ApJ...859..140C}, by \unit[0.16]{mag}, is that the error propagated from the \textit{Gaia} distance uncertainties is larger; \citet{2018ApJ...859..140C} placed tight constraints on the distance uncertainties included in their study. There are no signs of systematic zero-point differences among the various subtypes. The zero points were measured for a period of \unit[0.4]{day} (see the black dashed line in Fig.~\ref{fig:plr}). However, the difference between B- and other CB subtypes is greater than $\unit[3]{\sigma}$. This could also be an intrinsic characteristic of B-type systems, since they are not in thermal equilibrium. However, the significance of this result is compromised by small-number statistics, especially at the short-period end. Therefore, we will not include B-type CBs in our discussion.

We also compared our distance determinations, based on the \citet{2018ApJ...859..140C} PLRs, with the parallax-based distances of \citet{2018AJ....156...58B}. The mean difference between both measurements is $\langle \mathrm{DM}_\mathrm{PLR} - \mathrm{DM}_\varpi\rangle = \unit[0.012]{mag}$ ($\sigma=\unit[0.027]{mag}$), thus demonstrating the robustness of the CB PLR-based distance measurements. One possible explanation might be related to the intrinsic scatter in the $M$--$L$ and temperature--luminosity relations or that in the intrinsic properties, including the mass ratios, the orbital inclinations, and the fill-out factors (defining the extent to which a system's Roche lobe is filled). We explored the contributions of these three intrinsic parameters to the scatter to check whether the addition of a nonlinear component might be helpful to improve the accuracy of the PLRs.

\begin{figure*}[ht!]
\centering
\includegraphics[height=20cm]{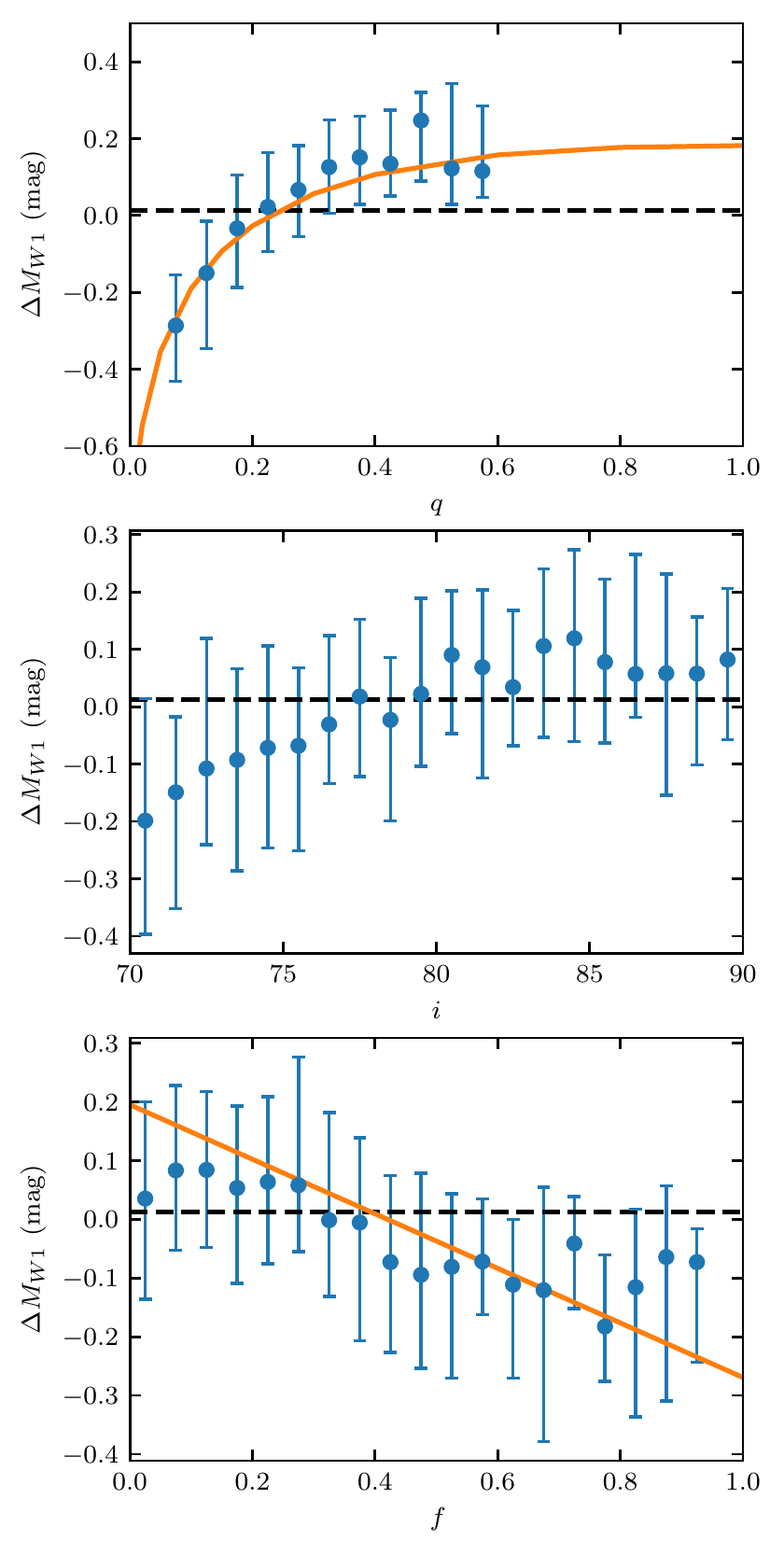}
\caption{Magnitude residuals ($\Delta M_{W1} = \mathrm{DM}_\mathrm{PLR} - \mathrm{DM}_\varpi = M_\varpi - M_\mathrm{DM}$) as a function of (top) mass ratio $q$, (middle) inclination $i$, and (bottom) fill-out factor $f$. The error bars represent the 25th and 75th percentiles of the distributions in each bin. The horizontal black dashed line is the mean magnitude difference $\langle \mathrm{DM}_\mathrm{PLR} - \mathrm{DM}_\varpi\rangle = \unit[0.036]{mag}$. Orange lines in the top and bottom panels represent theoretical models of the impact of changes in $q$ and $f$, respectively.\label{fig:zero}}
\end{figure*}

To construct Fig.~\ref{fig:zero}, we binned our sample into bins of different mass ratios $q$ (top), inclinations $i$ (middle), and fill-out factors $f$ (bottom), and we present the corresponding magnitude differences, $\Delta M_{W1} = \mathrm{DM}_\mathrm{PLR} - \mathrm{DM}_\varpi = M_\varpi - M_\mathrm{DM}$, in each bin. This latter parameter reflects the extent to which the luminosities are brighter than predicted by the PLRs; the smaller $\Delta M_{W1}$ is, the brighter the CBs are compared with the expected values. The error bars in Fig.~\ref{fig:zero} indicate the 25th and 75th percentiles of the distributions in each bin. The mean difference is shown as a vertical black dashed line. There are clear signs of local nonlinearities in the top and bottom panels, suggesting a dependence on the mass ratio and the fill-out factor. In the middle panel, $\Delta M_{W1}$ is consistent with the mean values in each inclination bin and remains flat.

In the top panel, a luminosity excess for $q < 0.2$ is obvious, which
could be explained by geometric differences of contact configurations
for different $q$ values. As the mass ratio decreases from unity to
zero, the radii of the primary and secondary Roche lobes will change
accordingly if the other parameters are fixed. According to
\citet[][their Tables 3-1 and 3-3]{1959cbs..book.....K}, the sum of
$r_\mathrm{1,R}$ (where `R' stands for `Roche lobe') and
$r_\mathrm{2,R}$ remains unchanged from $q=1$ to $q\approx 0.4$,
followed by a gentle increase toward lower values, thus leading to a
significant increase in the total surface area of the Roche lobe, $S
\propto r_\mathrm{1,R}^2 + r_\mathrm{2,R}^2$ for $q <
0.2$. Consequently, the total observed luminosities of our CBs
increase toward smaller $q$ values. We superimposed the theoretical
expectations for the effects of different $q$ values in the top panel,
which matches our results very well. This suggests a robust detection
of a $q$-induced zero-point shift in the PLR. A weak trend was
  also noticed for $i$, while $\Delta M_{W1}$ decreases for smaller
  inclination angles. However, this effect is not so significant
  compared with the size of the error bar. Similarly, the slight
decrease in $\Delta M_{W1}$ toward larger fill-out factors could be
related to the changes of the Roche lobes' surface areas. We used the
equations of \citet{2005ApJ...629.1055Y} to simulate this effect. The
adopted $q$ value is the sample's median mass ratio, $q=0.2$. As shown
in the bottom panel, as the fill-out factor becomes closer to unity,
the equivalent radii of the Roche lobes increase and render CBs with
larger fill-out factors brighter.

Based on Fig.~\ref{fig:zero}, we confirm that different intrinsic CB
parameters (in particular the mass ratio) have an impact on the PLR
zero points. The impact of other parameters is rather weak. The
influence of varying fill-out factors is relatively minor compared
with the effect of changing the mass ratios. This suggests that a
homogeneous CB sample, in terms of their mass ratios or fill-out
factors, might be helpful for future improvements of CB PLRs.

\section{Conclusions \label{sec:conclusions}}

In this paper, we have presented estimates of the fundamental
parameters of 2335 total-eclipsing CBs, based on a W--D-type
code. We used the $q$-search method to derive the mass ratios without
any knowledge of their radial velocity curves. The absolute parameters
were obtained by assuming that the primary stars of our sample CB
systems follow the ZAMS. A series of tests were designed to
  assess the accuracy and precision of our method. Our study has
shown the tremendous potential for statistical analysis of photometric
CB surveys. Our main results and conclusions are summarized below.

\begin{itemize}
\item Based on their masses and temperatures, our sample has been
  classified into three subtypes. It is composed of 1530 A-, 710 W-, and 95 B-type CBs.
\item The period distribution reveals that B-type CBs represent a different evolutionary phase compared with the other subtypes. A-type CBs have relatively longer periods than their W-type counterparts, tentatively suggesting that A-type systems may be less evolved.
\item The distribution of total CB masses, $M_\mathrm{tot}$, and mass
  ratios define a strip in phase-space. It has a well-defined edge at
  the lower $M_\mathrm{tot}$ limit. Although the majority of
    A-type CBs also lie in the strip, there is some hint suggesting a
    different distribution of A-type CBs. A large fraction of B-type
  CBs is located outside this strip.
\item It is likely that systematic differences in mass ratio and age exist between our large sample and other samples used previously. The latter was limited to the solar neighborhood.
\item There are no significant differences among the PLRs of A- and W-type CBs.
\item We confirm that the PLR zero-point deviates toward brighter magnitudes as the $q$ value decreases from $q=0.2$, which could be explained by geometric differences in the contact configurations for different $q$. This result may help us improve the accuracy of the PLRs in future studies.
\item An automated approach to deriving CB properties such as that employed here is a powerful tool for applications to future large samples. Combined with other information, such as the ages of star cluster hosts, the fundamental properties of CBs can be used to understand their evolution and death throes.
\end{itemize}

\acknowledgments L.D. and R.d.G. acknowledge research support from the National Natural Science Foundation of China through grants 11633005, 11473037, and U1631102. X.C. also acknowledges support from the National Natural Science Foundation of China through grant 11903045. The CSS is funded by NASA under grant NNG05GF22G issued through the Science Mission Directorate Near-Earth Objects Observations Program. The Catalina Real-Time Transient Survey is supported by the U.S.~National Science Foundation (NSF) under grants AST-0909182 and AST-1313422. This work has made use of data from the European Space Agency's (ESA) \textit{Gaia} mission (\url{https://www.cosmos.esa.int/gaia}), processed by the \textit{Gaia} Data Processing and Analysis Consortium (DPAC, \url{https://www.cosmos.esa.int/web/gaia/dpac/consortium}). Funding for DPAC has been provided by national institutions, in particular, those participating in the \textit{Gaia} Multilateral Agreement. This research was made possible through the use of the AAVSO Photometric All-Sky Survey (APASS), funded by the Robert Martin Ayers Science Fund and NSF AST-1412587. The Guoshoujing Telescope (LAMOST) is a National Major Scientific Project built by the Chinese Academy of Sciences. Funding for the project has been provided by the National Development and Reform Commission. LAMOST is operated and managed by the National Astronomical Observatories, Chinese Academy of Sciences.

\vspace{5mm}

\software{Wilson--Devinney program\, \citep{1971ApJ...166..605W}, Astropy \citep{2013A&A...558A..33A}, Matplotlib \citep{2007CSE.....9...90H}, dustmaps \citep{Green2018}}

\end{document}